\documentclass[twocolumn, astrosymb]{aastex631}

\usepackage{amsmath}
\usepackage{booktabs}

\newcommand{\ASL}{\color{black}}
\newcommand{\asl}{\color{black}}
\newcommand{\aref}{\color{black}}

\newcommand{\lya}{\ifmmode {\rm Ly\alpha} \else Ly$\alpha$\fi}
\newcommand{\ewlya}{\ifmmode W_{\rm Ly\alpha} \else $W_{\rm Ly\alpha}$\fi}
\newcommand{\ewhi}{\ifmmode W_{\rm HI} \else $W_{\rm HI}$\fi}

\newcommand{\fesclya}{\ifmmode f_{\rm esc}^{\rm Ly\alpha} \else $f_{\rm esc}^{\rm Ly\alpha}$\fi}
\newcommand{\fesclyc}{\ifmmode f_{\rm esc}^{\rm LyC} \else $f_{\rm esc}^{\rm LyC}$\fi}

\newcommand{\rsLya}{\ifmmode r_s^{\rm Ly\alpha} \else $r_s^{\rm Ly\alpha}$\fi}
\newcommand{\rsUV}{\ifmmode r_s^{\rm UV} \else $r_s^{\rm UV}$\fi}

\newcommand{\muv}{\ifmmode M_{\rm UV} \else $M_{\rm UV}$\fi}
\newcommand{\mstar}{\ifmmode M_{\star} \else $M_{\star}$\fi}
\newcommand{\msun}{\ifmmode M_{\odot} \else $M_{\odot}$\fi}

\received{2025 May 28}
\revised{2026 January 8}
\accepted{2026 January 10}

\shorttitle{LaCOS -- the connection between LyC Escape and Lyman-alpha Halos}
\shortauthors{Saldana-Lopez et al.}
\graphicspath{{./}{figures/}}

\begin{document}

\title{The Lyman-alpha and Continuum Origins Survey II: the connection between the escape of ionizing radiation and Lyman-alpha halos in star-forming galaxies}

\correspondingauthor{A. Saldana-Lopez}
\email{alberto.saldana-lopez@astro.su.se}

\author[0000-0001-8419-3062]{A. Saldana-Lopez}
\affiliation{Department of Astronomy, Oskar Klein Centre, Stockholm University, 106 91 Stockholm, Sweden}

\author[0000-0001-8587-218X]{M. J. Hayes}
\affiliation{Department of Astronomy, Oskar Klein Centre, Stockholm University, 106 91 Stockholm, Sweden}

\author[0000-0003-1767-6421]{A. Le Reste}
\affiliation{Minnesota Institute for Astrophysics, University of Minnesota, 116 Church Street SE, Minneapolis, MN 55455, USA}

\author[0000-0002-9136-8876]{C. Scarlata}
\affiliation{Minnesota Institute for Astrophysics, University of Minnesota, 116 Church Street SE, Minneapolis, MN 55455, USA}

\author[0000-0003-0470-8754]{J. Melinder}
\affiliation{Department of Astronomy, Oskar Klein Centre, Stockholm University, 106 91 Stockholm, Sweden}

\author[0000-0002-6586-4446]{A. Henry}
\affiliation{Center for Astrophysical Sciences, Department of Physics \& Astronomy, Johns Hopkins University, Baltimore, MD 21218, USA}
\affiliation{Space Telescope Science Institute, 3700 San Martin Drive Baltimore, MD 21218, USA}

\author[0000-0002-6085-5073]{F. Leclercq}
\affiliation{CNRS, Centre de Recherche Astrophysique de Lyon UMR5574, Univ Lyon, Ens de Lyon, F-69230 Saint-Genis-Laval, France}

\author[0000-0002-9613-9044]{T. Garel}
\affiliation{Observatoire de Gen\`eve, Universit\'e de Gen\`eve, Chemin Pegasi 51, 1290 Versoix, Switzerland}
\affiliation{CNRS, Centre de Recherche Astrophysique de Lyon UMR5574, Univ Lyon, Ens de Lyon, F-69230 Saint-Genis-Laval, France}

\author[0000-0001-5758-1000]{R. Amor\'in}
\affiliation{Instituto de Astrof\'isica de Andaluc\'ia (CSIC), Apartado 3004, 18080 Granada, Spain}

\author[0000-0002-7570-0824]{H. Atek}
\affiliation{Institut d'Astrophysique de Paris, CNRS, Sorbonne Universit\'e, 98bis Boulevard Arago, 75014, Paris, France}

\author[0000-0003-2722-8841]{O. Bait}
\affiliation{SKA Observatory, Jodrell Bank, Lower Withington, Macclesfield, SK11 9FT, UK}

\author[0000-0003-4166-2855]{C. A. Carr}
\affiliation{Center for Cosmology and Computational Astrophysics, Institute for Advanced Study in Physics \\ Zhejiang University, Hangzhou 310058,  China}
\affiliation{Institute of Astronomy, School of Physics, Zhejiang University, Hangzhou 310058,  China}

\author[0000-0002-0302-2577]{J. Chisholm}
\affiliation{Department of Astronomy, The University of Texas at Austin, 2515 Speedway, Stop C1400, Austin, TX 78712-1205, USA}

\author[0000-0002-0159-2613]{S. R. Flury}
\affiliation{Institute for Astronomy, University of Edinburgh, Royal Observatory, Edinburgh, EH9 3HJ, UK}

\author[0000-0001-6670-6370]{T. M. Heckman}
\affiliation{Department of Physics and Astronomy, Johns Hopkins University, 3400 North Charles Street, Baltimore, MD 21218, USA}

\author[0000-0002-6790-5125]{A. E. Jaskot}
\affiliation{Astronomy Department, Williams College, Williamstown, MA 01267, USA}

\author[0000-0003-1187-4240]{I. Jung}
\affiliation{Space Telescope Science Institute, 3700 San Martin Drive Baltimore, MD 21218, United States}

\author[0000-0001-7673-2257]{Z. Ji}
\affiliation{Steward Observatory, University of Arizona, 933 N. Cherry Avenue, Tucson, AZ 85721, USA}

\author[0000-0002-5235-7971]{L. Komarova}
\affiliation{Departament d'Astronomia i Astrof\'isica, Universitat de Val\'encia, C. Dr. Moliner 50, E-46100 Burjassot, Val\'encia, Spain}

\author[0000-0001-8792-3091]{Y-H. Lin}
\affiliation{Caltech/IPAC, 1200 E. California Blvd. Pasadena, CA 91125, USA}

\author[0000-0002-5808-1320]{M. S. Oey}
\affiliation{Astronomy Department, University of Michigan, 1085 South University Avenue, Ann Arbor, MI 48109, USA}

\author[0000-0002-3005-1349]{G. \"Ostlin}
\affiliation{Department of Astronomy, Oskar Klein Centre, Stockholm University, 106 91 Stockholm, Sweden}

\author[0000-0001-8940-6768]{L. Pentericci}
\affiliation{INAF-Osservatorio Astronomico di Roma, via Frascati 33, 00078, Monteporzio Catone, Italy}

\author[0000-0003-0470-8754]{A. Runnholm}
\affiliation{Department of Astronomy, Oskar Klein Centre, Stockholm University, 106 91 Stockholm, Sweden}

\author[0000-0001-7144-7182]{D. Schaerer}
\affiliation{Observatoire de Gen\`eve, Universit\'e de Gen\`eve, Chemin Pegasi 51, 1290 Versoix, Switzerland}
\affiliation{CNRS, IRAP, 14 Avenue E. Belin, 31400 Toulouse, France}

\author[0000-0001-5331-2030]{T. X. Thuan}
\affiliation{Astronomy Department, University of Virginia, P.O. Box 400325, Charlottesville, VA 22904-4325, USA}

\author[0000-0002-9217-7051]{X. Xu}
\affiliation{Department of Physics and Astronomy, Northwestern University, 2145 Sheridan Road, Evanston, IL, 60208, USA}
\affiliation{Center for Interdisciplinary Exploration and Research in Astrophysics (CIERA), 1800 Sherman Avenue, Evanston, IL, 60201, USA}



\begin{abstract}
One of the current challenges in galaxy evolution studies is to establish the mechanisms that govern the escape of ionizing radiation from galaxies. Here, we investigate the connection between Lyman Continuum (LyC) escape and the conditions of the Circumgalactic Medium (CGM), as probed by \lya\ halos (LAHs) in emission. We use \lya\ and UV continuum imaging data from the Lyman alpha and Continuum Origins Survey (LaCOS), targeting 42 nearby ($z \simeq 0.3$), star-forming galaxies with LyC observations (escape fractions of $\fesclyc \simeq 0.01-0.49$). LaCOS galaxies show extended \lya\ emission ubiquitously, with LyC emitters (LCEs) having more compact \lya\ morphologies than non-LCEs, and \lya\ spatial offsets that do not exceed the extent of the UV continuum. We model the diffuse LAHs using a combined S\'ersic plus exponential 2D profile, and find that the characteristic scale length of the \lya\ halo is ten times larger than the UV, on average. We unveil a significant anti-correlation between \fesclyc\ and the \lya\ Halo Fraction (HF, or contribution of the halo to the total \lya\ luminosity), which we propose as a new LyC indicator. Our observations show that halo scale lengths and HFs both scale positively with the optical depth of the neutral gas in the ISM, revealing a picture in which \lya\ and LyC photons in LCEs either emerge directly from the central starbursts or escape isotropically and, in the case of \lya, minimize the number of scattering interactions in a less-extended CGM. 
\end{abstract}



\section{Introduction}\label{sec:intro}

\begin{figure*}
    \centering
    \includegraphics[width=0.85\textwidth, page=1]{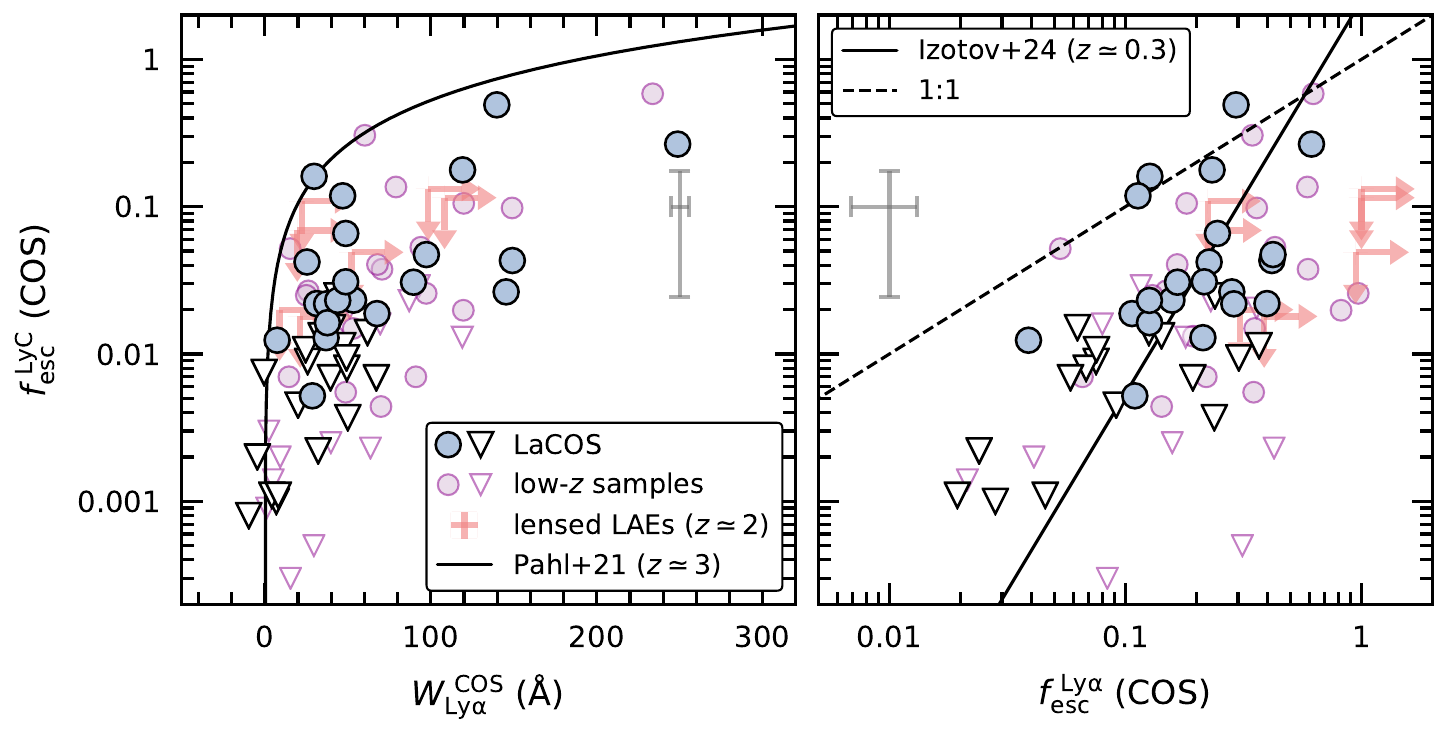}
\caption{{\bf LyC-to-\lya\ properties of the LaCOS sample.} Ionizing escape fraction (\fesclyc) versus the \lya\ equivalent width in the rest-frame (\ewlya, \emph{left}), and the \lya\ escape fraction (\fesclya, \emph{right}). Solid circles and downward triangles show LaCOS detections and upper limits, while shaded symbols in the background display other low-$z$ samples in the literature: \citet[][LzLCS]{Flury2022a}, \citet{Izotov2016a, Izotov2016b, Izotov2018a, Izotov2018b, Wang2019, Izotov2021} at $z \simeq 0.3$, and {\asl the lensed LAEs at $z \simeq 2.3$ by \citet[][pink arrows]{Citro2025}.} The solid lines draw the empirical relations from \citet{Pahl2021} at $z \simeq 3$ and \citet{Izotov2024} at $z \simeq 0.3$. {\asl The \ewlya\ and \fesclya\ are among the most robust one-dimensional \fesclyc\ indicators. Yet, the scatter in these relations is large, and the calibrations disagree between different redshifts}.}
\label{fig:lacos_sample}
\end{figure*}

Understanding the processes that caused the reionization of the intergalactic medium (IGM) around 1 billion years after the Big Bang is one of the current challenges in galaxy evolution theories \citep[e.g., reviews by][]{Barkana2001, Mesinger2016, Wise2019, Gnedin2022}. Constraining the shape and intensity of the cosmic ultraviolet background \citep[UVB, e.g.,][]{FG2009, HaardtMadau2012} during reinonization is essential, as it had significant impact over the thermal history of the IGM \citep[e.g.,][]{MiraldaEscude1994, Dayal2018} and, subsequently, in the formation and growth of baryonic structures \citep[e.g.,][]{Efstathiou1992, Gnedin2000, Okamoto2008}. Reionization also shaped the Cosmic Microwave Background (CMB) power spectrum, which in turn allows for a precise dating of when half of the IGM volume became ionized \citep[e.g.,][]{Planck2016}. However, the exact evolution of the neutral gas fraction in the IGM is still under debate. The timeline of reionization is inherently linked to the sources that produce and emit the necessary ionizing photons into the IGM, with bright but less numerous sources (whether galaxies or AGN) giving raise to a more rapid and late reonization \citep[e.g.,][]{MadauHaardt2015, Naidu2020}, while more numerous but faint counterparts leading to a more progressive and slow reionization process \citep[e.g.,][]{Robertson2013, Finkelstein2019, Rosdahl2022}. 

{\aref Measurements of the amount of HI ionizing (or Lyman Continuum, LyC; $\lambda_{\rm LyC} \leq 912$\AA) radiation that escape the sources during reionization are challenging \citep[e.g.,][]{Robertson2022}, mainly because of the increase in the IGM opacity at these high redshifts \citep[e.g.,][]{Inoue2014}}. Studies of LyC escape must be carried out in the nearby Universe to avoid absorption of the emergent ionizing photons by residual {\ASL neutral pockets} in the foreground IGM \citep[although see high-$z$ studies by e.g.,][]{Steidel2018, Fletcher2019, Begley2022, Saxena2022, RiveraThorsen2022}. As such, in the last decade the extragalactic community has embarked on a journey towards the discovery and characterization of the so-called analogs of cosmic reionizers \citep[e.g.,][]{Mascia2024}. For the first time, we have characterized the physical properties of LyC emitters \citep[e.g.,][]{Izotov2016a, Izotov2016b, Izotov2018a, Izotov2018b, Wang2019, Izotov2021, Flury2022a}. We have discovered that galaxies that emit significant amounts of LyC photons (called LyC emitters, hereafter LCEs) show overall young stellar populations (high H$\beta$ equivalent widths), a highly ionized medium (high [OIII]$\lambda5007$/[OII]$\lambda$3727,29), compact and intense star-formation (high SFR surface density), and a dust-poor (negative UV slopes) interstellar medium (ISM) with low column densities (weak absorption lines) of gas and metals \citep[e.g.,][]{Wang2021, Flury2022b, SaldanaLopez2022, Chisholm2022, Xu2023, Bait2024}. Detailed analysis of the stellar populations, ISM absorption and nebular emission line profiles \citep[e.g.,][]{Amorin2024, Carr2025} has revealed the importance of both radiative (stellar) and mechanical feedback (from supernovae) in ionizing and clearing out the channels needed for LyC photons to escape the ISM \citep[e.g.,][]{Flury2025}. 

Among the empirical \fesclyc\ relations, the ones involving the intensity and shape of the HI$\lambda1216$ spectral line (or \lya), are the most promising proxies \citep[e.g.,][]{Verhamme2017, Izotov2020}. According to idealized models \citep[e.g.,][]{Verhamme2015, Gronke2017, Garel2024} and simulations \citep[e.g.,][]{KakiichiGronke2021, Giovinazzo2024}, this is because these features, imprinted in the line profile via resonant radiative transfer, trace some of the properties of the surrounding neutral gas, such as column density or gas covering, which strongly regulate \lya\ and LyC escape \citep[e.g.,][]{Henry2015, Gazagnes2020}. Among these observables, the \lya\ equivalent width (\ewlya) and the escape fraction (\fesclya) stand out because of their applicability in high-$z$ systems \citep[e.g.,][]{Begley2024}. In Figure \ref{fig:lacos_sample}, we compile measurement of \ewlya\ and \fesclya\ as a function of the observed \fesclyc\ for samples of nearby galaxies \citep[see][and references therein]{Izotov2024}. The trends imply that galaxies with high LyC escape also show strong \lya\ with high \fesclya. However, the scatter on these one-dimensional \fesclyc\ relationships remains large, and observations of galaxies at higher redshift \citep[e.g.,][]{Pahl2021, Pahl2024, Kerutt2024} deviate from the local relations. {\ASL In fact, \citet{Citro2025} recently reported no LyC detection in a sample of strong, lensed LAEs with low dust contents, a lack that they attribute to the redshift evolution of the HI column density and dust content of the ISM of galaxies. These observations challenge previous interpretations based on local samples, suggesting that the extrapolation of $z \simeq 0$ \lya-based LyC estimators to the reionization epoch might not be fully correct \citep[see also the simulation work by][]{Maji2022, Choustikov2024}.}

The complicated 3D morphology of the ISM, the temporally varying star-formation, and the different timescales of the parameters involved in \fesclyc\ \citep[][]{Trebitsch2017, Mauerhofer2021}, presumably introduce significant scatter in the relations \citep[e.g.,][]{Choustikov2024a}. Therefore, unveiling the physics of LyC escape requires spatially resolved observations of the stars, gas and dust in the ISM of these galaxies, {\ASL so far missing for statistical LyC samples \citep[with the exceptions of the Sunburst Arc, Haro 11, Ion1 and J1316][]{RiveraThorsen2019, Komarova2024, Ji2025, MarquesChaves2024}.} With this goal in mind, in a previous paper we presented the \emph{Lyman-alpha and Continuum Origins Survey} (LaCOS), an HST imaging campaign targeting 42 nearby galaxies with LyC observations \citep{LeReste2025}, {\ASL a $z \leq 0.32$ sub-sample of the Low Redshift Lyman Continuum Survey \citep[LcLCS;][]{Flury2022a}}. In that work, we investigated the connection between the escape of ionizing photons and the \lya\ luminosity and equivalent width of the {\aref brightest UV-emitting star clusters.} 

In this paper, we aim to establish the link between the {\ASL properties of the extended \lya\ emission} and the physics of LyC escape, using LaCOS data. The manuscript is organized as follows. In Section \ref{sec:data}, we describe the LaCOS observations, data reduction and synthesis of the \lya\ maps. Section \ref{sec:results_nonparam} is devoted to basic morphological measurements of the \lya\ emission, such as sizes and UV-to-\lya\ offsets. In Section \ref{sec:results_modeling}, we model the LAHs in LaCOS, with special attention to UV and \lya\ characteristic scales. Section \ref{sec:discussion} discusses the connection between \fesclyc\ and LAHs, presenting a new indirect \fesclyc\ diagnostic based on the Halo Fraction (HF), i.e., the contribution of the \lya\ halo to the total \lya\ luminosity. We present a summary and our main conclusions in Section \ref{sec:conclusions}. 

Throughout, we use a flat cosmology with $\{H_0, \Omega_M, \Omega_{\Lambda}\} = \{70~{\rm km~s^{-1}~Mpc^{-1}}, 0.3, 0.7\}$ {\aref and distances are reported in physical (kpc) units.} The AB magnitude system \citep{OkeGun1983} is adopted. We use the survival Kendall $\tau$ correlation test \citep{ATS1996} to assess the degree of correlation between variables, using the scheme developed in \citet{Isobe1986} that allows for the inclusion of censored data. We use the code developed in \citet{Herenz2025}\footnote{The \textsc{kendall} code \citep{Herenz2025} is publicly available on \url{https://github.com/Knusper/kendall}.} adapted from \citet{Flury2022b}. Following the LzLCS sample convention, we deem a correlation ($\tau>0$) or anti-correlation ($\tau<0$) significant when $p_{\rm val.} \leq 1.35 \times 10^{-3}$ ($3\sigma$ confidence). In other words, we reject the null hypothesis over this threshold. We will also define as marginal or tentative those correlations showing $1.350 \times 10^{-3} \leq p_{\rm val.} \leq 2.275 \times 10^{-2}$ (2 to $3\sigma$ significance). 

\begin{figure*}
    \centering
    \includegraphics[width=0.99\textwidth, page=1]{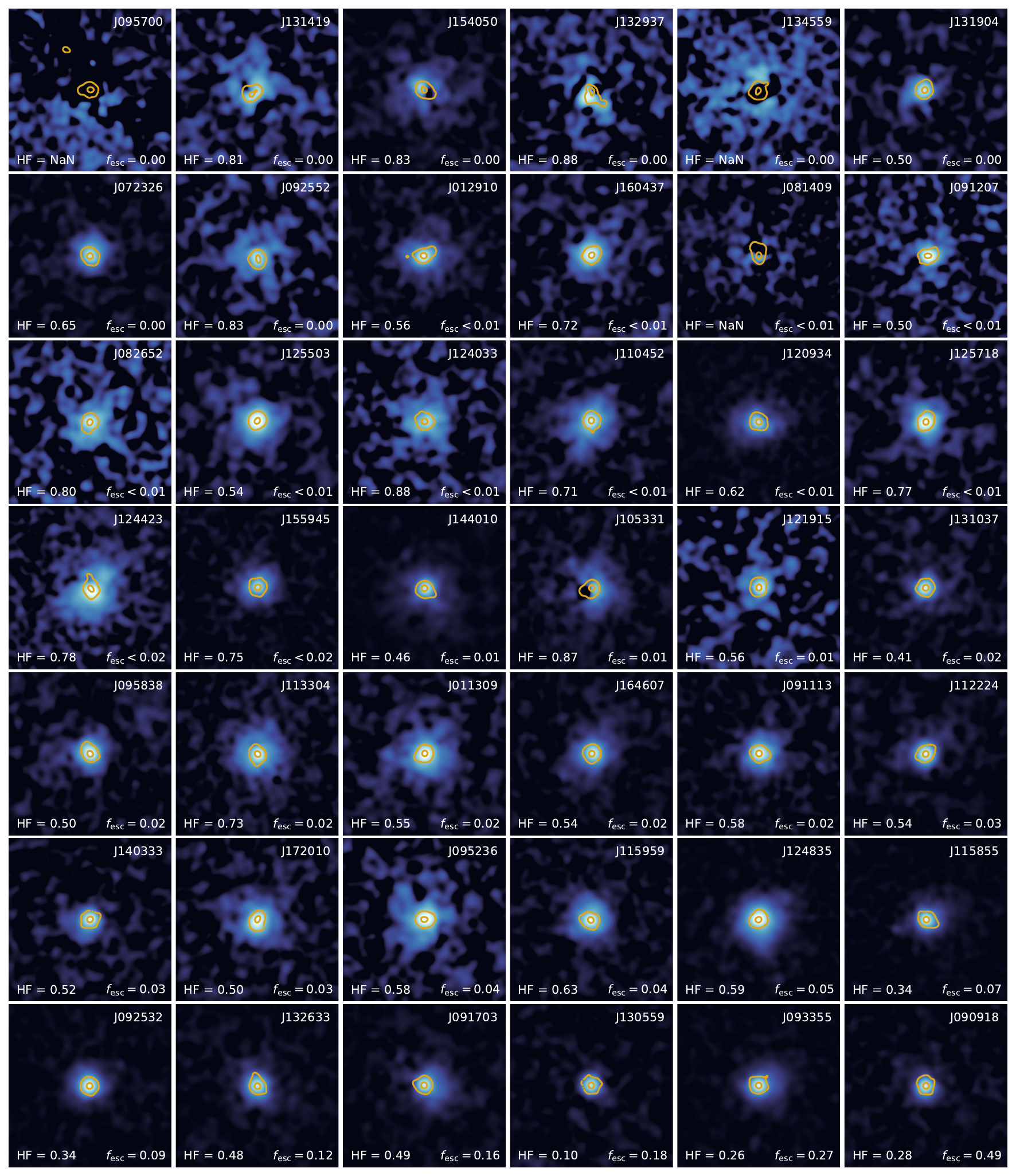}
\caption{{\bf Extended \lya\ emission in LaCOS galaxies (${\rm 15~kpc \times 15~kpc}$ cutouts).} A {\asl smoothed version of the \lya\ emission is shown in blue (${\rm arcsinh}$ scale), with the orange contours depicting the more compact, UV continuum counterpart (applying a 5~pix Gaussian filter).} White labels show the galaxy ID, measured LyC escape fraction for every object (\fesclyc, including upper limits), and the estimated \lya\ Halo Fraction (HF) (see Sect.\,\ref{sec:results_modeling}). Panels are sorted by ascending \fesclyc.}
\label{fig:lacos_lahs_rgb}
\end{figure*}

\section{The Lyman Alpha and Continuum Origins Survey (LaCOS)}\label{sec:data}
The Lyman-Alpha and Continuum Origins Survey \citep[LaCOS -- ID GO17069; PIs Hayes, Scarlata, see][]{LeReste2025}, was built from the LzLCS survey \citep{Flury2022a}, the largest sample of nearby galaxies with ionizing continuum observations. The LzLCS sample comprised 66 galaxies at $z \simeq 0.3$ from the Sloan Digital Spectroscopic Survey Data Release 17 \citep[SDSS-DR17,][]{Blanton2017_SDSSdr17}, with available observations from the Galaxy Evolution Explorer \citep[GALEX,][]{Morrissey2007_GALEX}. These galaxies were selected to have either high $O_{32}$ ($>3$), high $\Sigma_{\rm SFR}$ ($>0.1~{\rm \msun yr^{-1} kpc^{-2}}$), or blue UV colors ($\beta_{\rm UV}<-2$), properties thought to primarily influence LyC escape. {\ASL AGN and composite systems were excluded from the final sample using classical BPT emission line diagnostics \citep[e.g.,][]{BPT1981}.}

All LzLCS galaxies were observed with HST/COS using the G140L grating, probing the LyC window ($850-900$\AA) at the redshift of the observations ($z=0.22-0.45$). In order to constrain the intrinsic production of ionizing continuum photons, the FUV stellar continuum redder than 912\AA\ was modeled via spectral fitting \citep{SaldanaLopez2022}. Together with the observed LyC fluxes from COS, the fiducial \fesclyc\ is then estimated dividing the former by the latter \citep{Flury2022b}, resulting in 35 LyC detections at $2\sigma$ significance and $\fesclyc = 0.01 - 0.49$, and 31 non-detections with $1\sigma$ upper limits in \fesclyc\ of $\leq 1\%$, typically. 

The main goal of LaCOS is to spatially map the emission of \lya\ radiation, and use the \lya\ intensity and morphology as diagnostics of LyC escape. To do so, LaCOS employs the effective narrow band technique \citep{Hayes2009}, which allows the construction of emission line maps by using two nested long-pass filters of the \emph{Solar Blind Channel} (SBC) onboard of HST. With the bluer filter sampling emission line plus stellar continuum, and the redder filter sampling continuum only, the emission line map is obtained by scaling the intensity measured in the redder filter and subtracting it from the bluer image. {\ASL LaCOS is the $z \leq 0.32$ subample of the LzLCS, corresponding to the redshift range allowing the imaging of \lya\ using the SBC/F150LP and F165LP ramp filters.} At higher-$z$, the \lya\ line redshifts into the reddest bandpass on SBC. Following this criterion, we select 41 out of 66 LzLCS galaxies. One additional galaxy from the literature with available archival imaging was added \citep{Izotov2016a}, resulting in a sample of {\ASL 42 representative galaxies for the LaCOS survey.} 

As shown in \citet{LeReste2025}, the distribution of physical properties of this sub-sample is similar to that of the parent LzLCS survey, with absolute UV magnitudes of $-21 \leq \muv \leq -18$, stellar masses and SFRs in the range $\log \mstar/\msun = 7.5-10.5$ and ${\rm SFR / \msun yr^{-1}} = 1-30$, gas-phase metallicities $12 + \log {\rm O/H} = 7.5-8.5$, Balmer-line strengths ($W_{\rm H\beta}$) up to 300\AA\ and UV colors of $\beta_{\rm UV} = -2.6$ to $0.3$. In Fig.\,\ref{fig:lacos_sample} we compare the \lya\ and LyC properties of LaCOS to the parent LzLCS sample \citep{Flury2022b} and other measurements from the literature \citep{Izotov2016a, Izotov2016b, Izotov2018a, Izotov2018b, Wang2019, Izotov2021, Citro2025} that include \lya\ and LyC information. As mentioned in the Introduction, both \ewlya\ and \fesclya\ correlate with \fesclyc\ \citep[see relations by][]{Pahl2021, Izotov2024}, although the scatter is substantial. Quantifying this scatter is one of the main scientific objectives of LaCOS. 

The 41 LaCOS galaxies were observed following a five-band imaging strategy with HST, using the \emph{Advanced Camera for Surveys} (ACS, 2~orbits/target) and the \emph{Wide Field Camera 3} (WFC3, 1~orbit/target). The ACS/SBC F150LP and F165LP filters captured the \lya\ emission and rest-UV continuum, while the WFC3/UVIS F438W, F547M and F850LP filters probed the Balmer break and rest-optical continuum, respectively. The combination of long-pass filters in the UV and medium and broad-band filters in the visible, {\ASL allows for a spatially resolved study of the diffuse gas in the ISM and the Cirumgalactic Medium (CGM) probed by \lya\ emission,} of the spatial distribution of the young star-forming regions, as well as of the older stellar populations and dust extinction. 

The SBC and UVIS images were reduced following the methods described in \citet{LeReste2025}, with custom routines to mitigate the effect of dark current over the SBC frames, perform the rejection of cosmic rays from the UVIS files \citep[following][]{vanDokkum2001}, and an additional background subtraction applied to all data frames. Similar methods have been used in the E/LARS survey \citep[e.g.,][]{Ostlin2014, Melinder2023} and other studies at the same redshift as this work \citep{Runnholm2023}. Individual frames for each filter were then registered and co-added together to the native UVIS pixel scale of $0.04''$, resulting in a total exposure time of around $2,000$s and $2,500$s for the SBC frames, and 500, 620 and 700s for the UVIS filters (in ascending order of central wavelength). Finally, all images were convolved to a common Point Spread Function (PSF), which is constructed from all of the SBC and UVIS filters to be the broadest PSF at any given radius, following the methods in \citet{Melinder2023}. The final LaCOS images have a $0.1''$ PSF Full Width at Half Maximum (FWHM), corresponding to a physical scale of 360pc at the median redshift of $z\simeq 0.27$, and effectively probing sub-kpc scales in the ISM and CGM. All frames were corrected for the Milky Way extinction using the \citet{Fitzpatrick1999} extinction law and measurements of Galactic $E_{\rm B-V}$ from \citet{Green2018}. {\aref The data products for the HST observation of LaCOS galaxies (including archival data) are being released at the Barbara A. Mikulski Archive for Space Telescopes (MAST) as a High Level Science Product\footnote{Also accesible via \url{https://archive.stsci.edu/hlsp/lacos}}, with the following
DOI: \href{https://archive.stsci.edu/doi/resolve/resolve.html?doi=10.17909/j4qd-ev76}{10.17909/j4qd-ev76}.}

As described in \citet{LeReste2025}, spatially resolved maps of \lya\ were constructed by matching the equivalent width of \lya\ (\ewlya) measured over the COS/G140L spectra \citep{Flury2022a}, to the ones obtained within a $2.''5$-aperture in the F150LP and F165LP PSF convolved frames. We refer to the former paper for a thorough overview of the data reduction process and creation of the \lya\ maps for the LaCOS survey. Figure \ref{fig:lacos_lahs_rgb} shows color composites of the (smoothed) \lya\ maps, with the intensity of the UV continuum overlaid. The {\ASL significant} detection of \lya\ emission extending beyond the UV starlight suggests the presence of \lya\ halos (LAHs) in most, if not all, LaCOS galaxies, as we will discuss in detail in the forthcoming sections. 

\begin{figure}
    \centering
    \includegraphics[width=0.99\columnwidth, page=1]{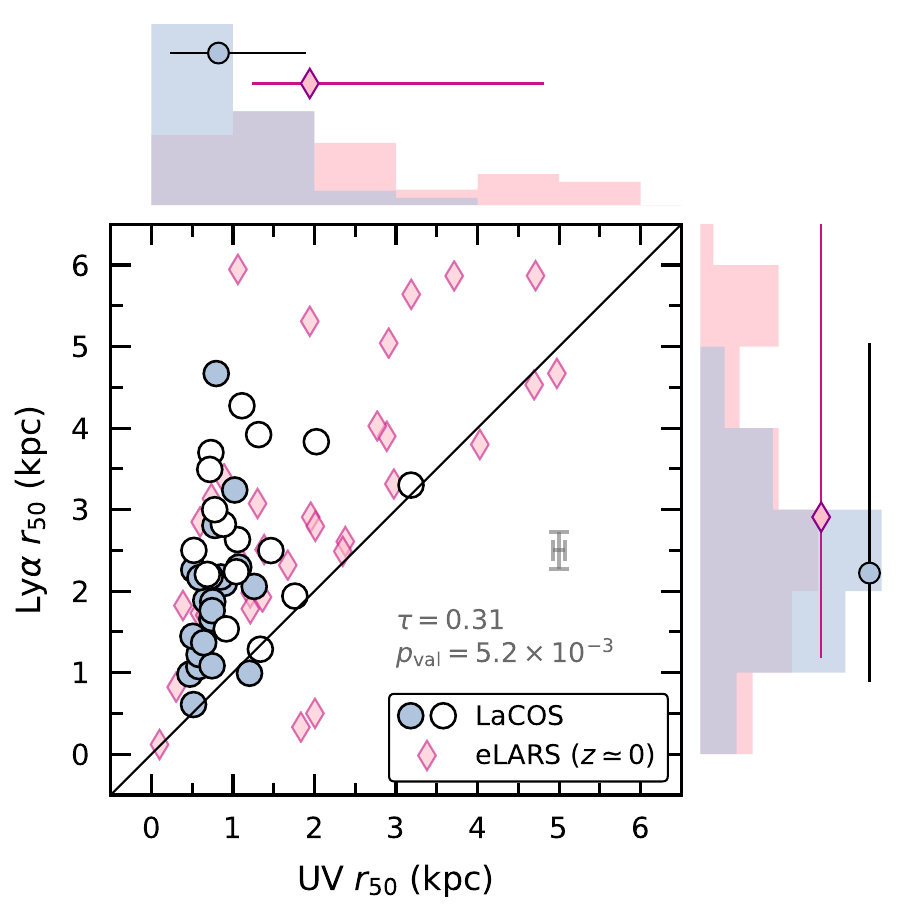}
\caption{{\bf Comparison between the extent of \lya\ and UV emission in LaCOS,} as measured by the half-light radius ($r_{50}$). {\aref Filled and open circles show LaCOS LyC detections and upper limits, respectively.} For reference, similar measurements from the eLARS survey are plotted in pink diamonds \citep{Melinder2023}, with the solid line indicating the one-to-one relation. {\aref Histograms show the size distributions projected on each axis, with the error bars encompassing the interquartile range.} {\ASL Error bars represent the characteristic (median) uncertainty on each axis.} With respect to their UV counterpart, LaCOS galaxies show extended \lya\ emission almost ubiquitously.}
\label{fig:LyaUV_size}
\end{figure}

\section{Compact \lya\ emission around Lyman Continuum Emitters}\label{sec:results_nonparam}
{\asl Several astrophysical phenomena can contribute to the presence of \lya\ emission in the CGM of galaxies: \lya\ cooling radiation produced by inflowing gas \citep[e.g.,][]{Dijkstra2009, FaucherGiguere2010}, scattering of nebular and continuum \lya\ photons emitted from the star-forming regions \citep[e.g.,][]{Laursen2009, Zheng2010, Dijkstra2012}, the production of \lya\ photon \emph{in-situ} via recombination of ionizing radiation \citep[e.g.,][]{Cantalupo2005, Furlanetto2005, Martin2015, MasRibas2016, Carr2021}, or direct \lya\ emission from unresolved galaxy satellites within the same dark-matter halo \citep[e.g.,][]{Shimizu2010, MasRibas2017}.} Because of the need for spatially resolved observations at multiple wavebands (e.g., \lya, UV, H${\rm \alpha}$), and the low-surface brightness of some of the targeted features, disentangling the different scenarios is challenging \citep[e.g.,][]{Bridge2018}. 

\begin{figure*}
    \centering
    \includegraphics[width=0.85\textwidth, page=1]{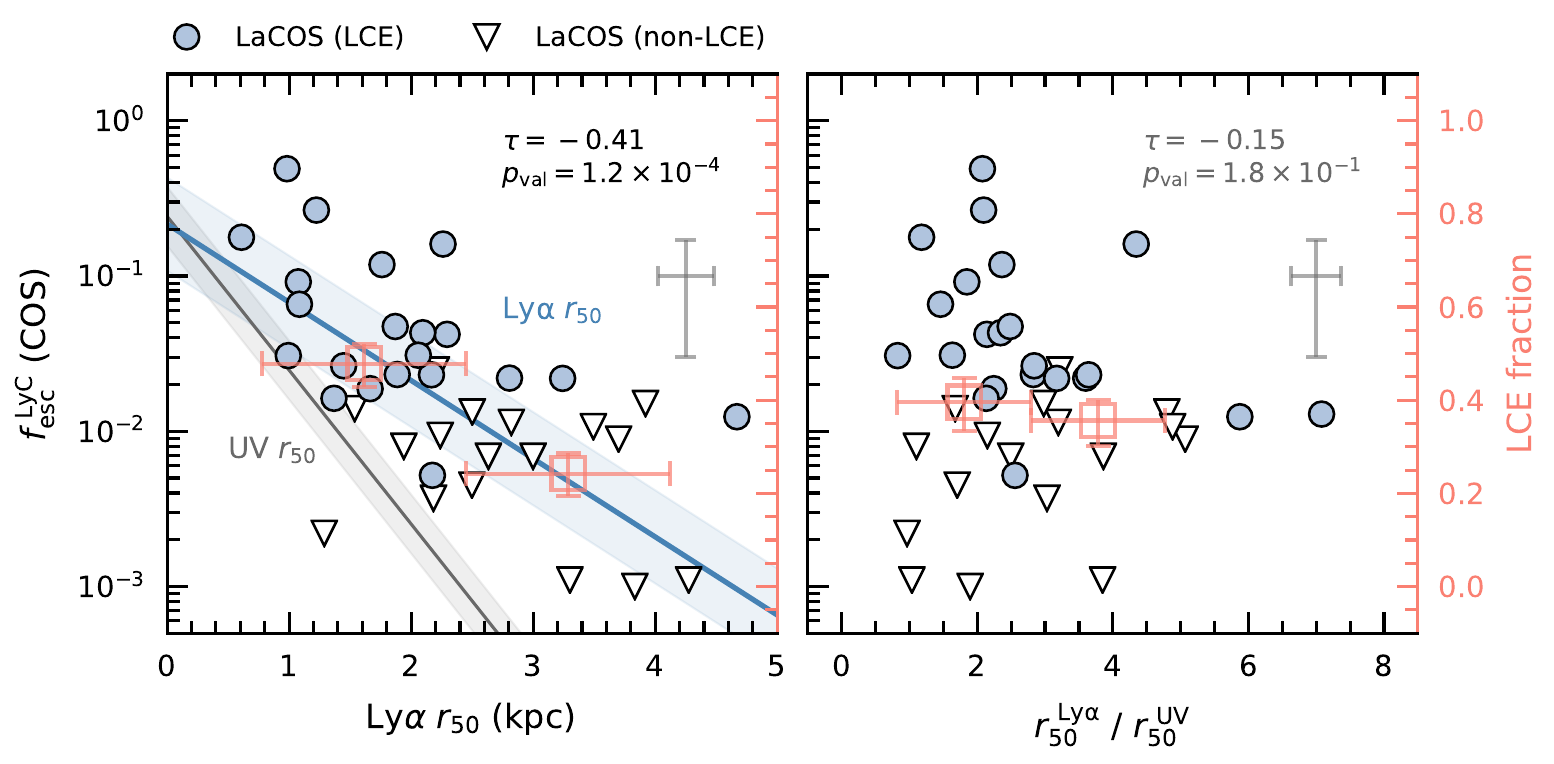}
\caption{{\bf The relation between the ionizing escape fraction (\fesclyc) and the extent of the \lya\ emission (\emph{left}), and the \lya-to-UV size ratio (\emph{right}).} {\aref Filled circles and downward triangles show LaCOS LyC detections and upper limits, respectively. The LCE detection fraction is also shown through squared open symbols in the right vertical axis. The results from the survival Kendall correlation test, including censored data, can be found in the inset. Linear fits to the decline in \fesclyc\ with the size of both the UV continuum and \lya\ are plotted in blue and gray lines (Eq.\,\ref{eq:fesc_rUV} and \ref{eq:fesc_rLya}). While these linear fits indicate that LCEs may have more compact \lya\ than non-LCEs respect to the UV continuum (\emph{left}), individual data points do not reflect this behavior (\emph{right}), highlighting the lack of ability of simple size measurements to fully reproduce the morphology of the \lya\ emission, and the need of a more sophisticated modeling (see Sect.\,\ref{sec:results_modeling}).}}
\label{fig:fesc_Lya}
\end{figure*}

However, models predict scattering to be the main contributor to \lya\ {\ASL outside of star-forming regions but within the innermost CGM} \citep[projected distances of $\leq 10{\rm ~kpc}$;][]{Lake2015, Byrohl2021, Mitchell2021}, which is easily studied with LaCOS imaging (see Fig.\,\ref{fig:lacos_lahs_rgb}). In this situation, star-forming regions copiously produce \lya\ radiation \citep[e.g.,][]{Schaerer2003} that, because of its high cross section {\asl and the large abundance of hydrogen} \citep[e.g.,][]{Neufeld1990}, can resonantly scatter in the gaseous halo creating a diffuse emission beyond the location of the UV sources. To characterize the morphology of the extended \lya\ emission around LaCOS galaxies, we start by comparing the light distribution of the \lya\ and the UV continuum. 

\subsection{The extent of the \lya\ and UV emission}
First, we compute the radial intensity profile for both the UV continuum and \lya\ images of each source. This is done by measuring the total flux encompassed within concentric circular apertures in radial increments of 2 pixels, up to 200 pixels in total, starting from the centroid of the UV continuum band. Then, we read out the radii at which 20, 50 and 90 per cent of the flux within the 200 pixel circle is contained, getting $r_{20}, r_{50}$ and $r_{90}$, respectively {\ASL (see Table \ref{tab:LyaUV_analysis})}. Uncertainties on these measurements ($1\sigma$) are reported by Monte Carlo sampling the individual pixels in the UV and \lya\ images with the corresponding error frames. 

{\aref First, the use of circular apertures is justified by the visual symmetry and compact morphology of the \lya\ intensity maps: while perhaps over-simplistic, the circular annuli capture the amount of light within fixed radius with no dependency on the clumpy underlying morphology. Furthermore, this approach allows for direct comparison with existing measurements at high-redshift.  Second, the choice of a large, 200 pixel-wide aperture (or $8''$) is made so that it contains the total \lya\ flux even for the largest galaxies in the sample (i.e., J081409, J095700, J134559). This aperture size corresponds to $\simeq 35{\rm ~kpc}$ in diameter for the LaCOS median redshift of $z=0.27$, a scale that remains almost invariant across the sample, because of the similar redshifts of the LaCOS galaxies ($0.22\leq z\leq 0.32$, or a $\simeq 20$\% variation in kpc units).}

{\aref Finally, for a proper comparison between the extent of the \lya\ and the UV emission, a similar limiting depth for both observations is required. In LaCOS, this is deliberately achieved by assigning comparable integration times\footnote{{\aref With the exception of J092532 and J124835, for which we used additional archival observations, significantly increasing the length of the F165LP exposures \citep[see][]{LeReste2025}.}} to the F150LP and F165LP exposures. For extended sources with the same exposure time, the limiting surface brightness will depend on the redshift as $(1+z)^{-4}$ \citep{Giavalisco1996}. Once again, due to the narrow redshift range covered by the LaCOS galaxies, the reached surface brightness limit will not vary much between sources ($\simeq 25\%$ across the sample). We compute the surface brightness limit for the UV and \lya\ images of each source, by measuring the average value of the standard deviation of the flux (from the weight maps) over a 5-pixel-wide annulus of 200 pixel size. The resulting limiting depths for the UV and \lya\ are comparable ($\simeq 2 \times 10^{-17}{\rm ~erg~s^{-1}~cm^{-2}~\AA^{-1}~arcsec^{-2}}$), where the \lya\ flux was divided by the 109.2\AA\ band-pass of the F150LP filter.}

{\aref Thus, we measure \lya\ radii in the range $r_{50} = 0.6-7.7{\rm ~kpc}$, ($r_{20} = 0.3-1.8{\rm ~kpc}, r_{90} = 3.6-23.7{\rm ~kpc}$), while the for the UV continuum we measure $r_{50} = 0.5-3.2{\rm ~kpc}$ ($r_{20} = 0.2-1.5{\rm ~kpc}, r_{90} = 2.6-20.5{\rm ~kpc}$). We note that the UV half-light radii are slightly above the measurements reported in \citet{Flury2022a} from the COS acquisition images, due to the lower surface brightness and larger field of view reached by our SBC observations compared to COS \citep[see also][]{LeReste2025}.} 

In Figure \ref{fig:LyaUV_size} we compare the 50\%-light radius of \lya\ and UV emissions, {\aref resulting in a tentative correlation between the two}. LaCOS galaxies show a diversity of \lya\ to UV sizes, having $r_{50}^{\rm \lya} / r_{50}^{\rm UV} = 0.8 - 7.0$, with a mean of 2.8, in agreement with the median of 2.9 reported by \citet{Hayes2013} in local galaxies. {\asl Although in a photo-ionization scenario, the gas (and therefore the \lya\ emission) is expected to extend further than the stars,} a simple calculation of the \lya\ and UV continuum surface brightnesses (by summing up the flux in concentric circular annuli instead of apertures) reveals faint \lya\ emission ($2\sigma$ detection) at distances as far as {\ASL 10 times from the edge of the UV continuum,} with a median of 4.5 times. 

Altogether, this confirms that the emergent \lya\ emission is significantly more extended than the UV for the vast majority of LaCOS galaxies, and extends over scales corresponding to the inner CGM domain \citep[$5-50{\rm ~kpc}$, e.g.,][]{Tumlison2017}. When compared to other measurements of extended \lya\ emission around low-$z$ LAEs, such as the eLARS\footnote{Extended Lyman-Alpha Reference Sample (eLARS)} galaxies \citep[e.g.,][]{Melinder2023}, our $r_{50}^{\rm \lya} / r_{50}^{\rm UV}$ ratios, although roughly compatible, lay in the lower bound region of the parameter space, showing smaller \lya\ and UV size than the average eLARS galaxy. This is by selection, as the eLARS survey added some large, nearby galaxies to the original LARS starburst galaxy sample \citep{Ostlin2014}. {\ASL At higher redshifts ($2.8 \leq z \lesssim 6$),} \citet{Claeyssens2022} reported higher \lya\ to UV $r_{50}$ ratios than this work (4.8 and 12, on average). 

\begin{figure*}
    \centering
    \includegraphics[width=0.95\textwidth, page=1]{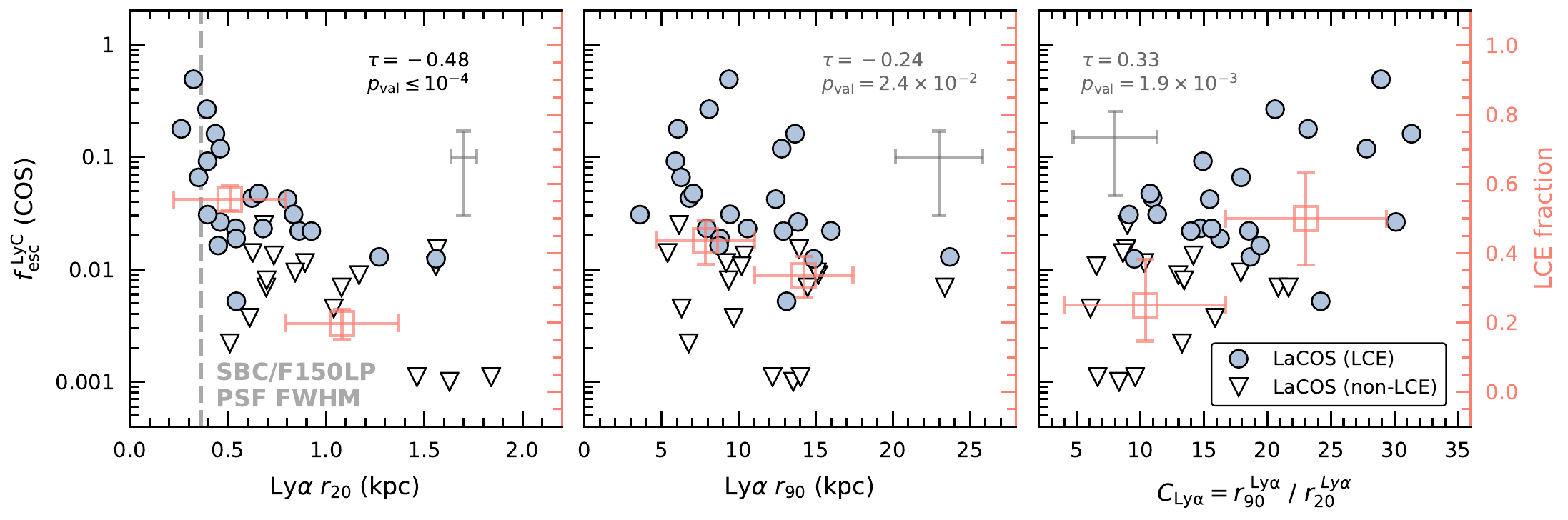}
\caption{\asl {\bf Ionizing escape fraction (\fesclyc) versus the 20\%- and 90\%-light radius, and the concentration of the \lya\ emission, defined as $C_{\rm \lya} = r_{90}^{\rm \lya}/r_{20}^{\rm \lya}$.} The LCE fraction (tentatively) increases towards more compact galaxies in \lya, due to the underlying correlation between \fesclyc\ and \lya\ $r_{20}$.}
\label{fig:fesc_CLya}
\end{figure*}

Figure \ref{fig:fesc_Lya} (left) shows the escape fraction of ionizing photons (\fesclyc) as a function of the \lya\ half-light radius ($r_{50}$) for the LaCOS survey. In order to assess the significance of the correlation, we perform a survival Kendall correlation test \citep{ATS1996}, which properly accounts for \fesclyc\ upper limits in the ranking. Our Kendall test reveals a strong and significant anti-correlation between \fesclyc\ and \lya\ $r_{50}$, meaning that galaxies with smaller \lya\ radii tend to have higher escape fractions. In the same panel, we also plot the LCE fraction as the number of LyC detections over the total, by splitting the sample into equally populated $r_{50}^{\lya}$ bins via the median value. Similarly, the LCE fraction increases from $0.24 \pm 0.04$ at $r_{50}^{\lya} = 3.2{\rm ~kpc}$ to $0.48 \pm 0.05$ at $r_{50}^{\lya} = 1.6{\rm ~kpc}$. {\ASL These LCE fractions were calculated using the methods in \citet{Flury2022a}\footnote{\url{https://github.com/sflury/histogram}}, in which each fractional bin is Poisson binomial and representative of the independent sampling of each datum from its respective normal distribution \citep[characterized by its uncertainty, otherwise see][]{Gehrels1986}.}

{\asl If the extended \lya\ emission is produced via scattering within the gaseous halo, a connection between \fesclyc\ and the \lya\ size is expected. The relation shown in Fig.\,\ref{fig:fesc_Lya}, however,} {\aref may be affected by the already reported trend between the escape fraction and the size of the star-forming regions traced by the UV half-light radius \citep{LeReste2025}, and the underlying \lya-to-UV correlation itself (Fig.\,\ref{fig:LyaUV_size}).} The emergence of the \fesclyc\ to UV-size relation is attributed to the influence of feedback in LyC escape \citep[][]{Kimm2017, Trebitsch2017}. {\asl The affinity of strong LCEs for high $\Sigma_{\rm SFR}$ and low UV size \citep{Naidu2020, Flury2022b} suggests that higher concentrations of star formation provide the mechanical feedback necessary to clear LyC escape paths \citep[][]{Bait2024, Bait2025, Flury2025}. Alternatively, the low metallicity of these young starbursts may delay the explosion of supernovae \citep{JecmenOey2023}, requiring the ionizing feedback to contribute, which will be more efficient at ionizing the surroundings in more compact systems \citep[e.g.,][]{Jaskot2019, Carr2025}.} {\ASL Consistently, \citet{Amorin2024} and \citet{Komarova2025} have shown that high ionized gas velocities are preferentially found in stronger leakers.}

{\asl To circumvent the aforementioned bias, in Fig.\,\ref{fig:fesc_Lya}} we perform log-linear fits to the \fesclyc\ versus the UV and \lya\ $r_{50}$ separately, using the \textsc{linmix}\footnote{{\aref \textsc{linmix} is a Python-based Bayesian fitting code that allows the user to model a two-dimensional data set with a linear regression, accounting for errors on both variables and intrinsic random scatter, with the capability of including censored (upper or lower limits) data. A python version of \textsc{linmix} can be found in \url{https://linmix.readthedocs.io/en/latest/index.html}.}} Bayesian fitting code \citep{linmix}. {\aref For upper-limits in the escape fraction, \textsc{linmix} marginalizes over the unobserved true values conditional on being below the observed limit.} We obtain, 
\begin{equation}
    \log \fesclyc = (-0.61 \pm 0.17) \cdot r_{50}^{\rm UV} - (0.98 \pm 0.19)
    \label{eq:fesc_rUV}
\end{equation}

\noindent and
\begin{equation}
    \log \fesclyc = (-0.50 \pm 0.13) \cdot r_{50}^{\rm \lya} - (0.67 \pm 0.30)
    \label{eq:fesc_rLya}
\end{equation}

\noindent respectively. {\ASL Given that the slope of the \fesclyc\ to UV-size relation is steeper than the \lya\ fit, this suggest that \lya\ to UV size ratio may decrease with increasing \fesclyc. 

{\aref Consistently, \citet{Leclercq2024} recently found that strong LCEs appear uniformly compact in the low ionization gas-phase (traced by MgII$\lambda\lambda2796,2803$ and [OII] emission) with respect to the UV starlight, suggesting they do not have extended neutral gas halos. The non-LCEs, on the other hand, showed a diversity of low-ionized gas configurations. In the right panel of Figure \ref{fig:fesc_Lya}, we further stress this hypothesis by plotting the \lya-to-UV size ratio as a function of \fesclyc. While the linear fits in the left panel indicate that LCEs may have more compact \lya\ than non-LCEs with respect to the UV continuum, individual data points do not reflect this behavior.} 

{\aref We note that, while the circular apertures adopted here may not capture azimuthal variations in the light profile, they are firstly a direct and non-parametric way to capture the radial light profile (growth of the integrated flux with radius), and secondly they can easily be adopted for galaxies in the high-redshift Universe, where the average S/N per pixel may be lower and parametric fitting methods may fail. Moving forward these limitations, in Sect.\,\ref{sec:results_modeling} we introduce a more sophisticated modeling to describe the morphology of \lya\ in the LaCOS sample.}

\subsection{The morphology of extended \lya\ emission}
{\ASL To gain more insights into the connection between the properties of the extended \lya\ emission and \fesclyc, we now characterize the morphology of the \lya\ images according to the concentration parameter \citep{Conselice2003}, defined in this work as $C_{\rm \lya} = r_{90}^{\rm \lya}/r_{20}^{\rm \lya}$. We observe a wide range of $C_{\rm \lya}$ values in LaCOS, with $r_{90}$ being between five and 30 times larger than \lya\ $r_{20}$. The LyC escape fraction (\fesclyc) is plotted against the same $C_{\rm \lya}$ statistic in Figure \ref{fig:fesc_CLya}, {\aref alongside the \lya\ $r_{20}$ and $r_{90}$ measurements for the same sample}. 

We find that both \fesclyc\ and the LCE fraction tentatively increase with the $C_{\rm \lya}$ parameter. In other words, LCEs and galaxies with high \fesclyc\ tend to show more concentrated \lya\ light distributions, in line with the results hinted in the previous section. This tentative, positive correlation between \fesclyc\ and $r_{90}^{\rm \lya}/r_{20}^{\rm \lya}$ is driven by the underlying, strong anti-correlation between \fesclyc\ and $r_{20}$, while there is no correlation at all between \fesclyc\ and $r_{90}$. Once again, this illustrates the lack of ability of simple size measurements to reproduce the morphology of \lya\ in compact galaxies. {\ASL Even though PSF effects are not accounted in this simple size analysis, we note that most of the LaCOS galaxies fall above the $C_{\rm \lya} \simeq 7.3$ value expected from an exponential light profile. This suggests that more complicated functional forms, specifically including steeper profiles at shorter radii, are needed to reproduce the full \lya\ light profile (see Sect.\,\ref{sec:results_modeling}).}} 

\subsection{\lya\ to UV continuum offsets}
In this section, we calculate the spatial offset between the centroids
of the \lya\ and UV emission ($\Delta_{\rm \lya - UV}$). These offsets may indicate whether \lya\ photons are produced in or scattered away from star-forming regions responsible for the UV, supporting one of the aforementioned scenarios for extended \lya\ \citep[e.g.,][]{Bhagwat2025}. {\ASL For instance, small offsets could indicate star formation knots off-centered from the main body of the starburst \citep[as in the remarkable case of Haro 11][]{Komarova2024}, while larger ones may favor satellite galaxy emission.} 

For around half of the LaCOS sample, the estimated offsets {\ASL using the \texttt{photutils.centroids} routine \citep{photutils}} are larger than half of the PSF FWHM of our SBC observations. For those, we measure $\Delta_{\rm \lya - UV}$ ranging from 0.14 to 4.31 kpc, with a mean of $1.63{\rm ~kpc}$. Typical values found in the $2 \leq z \leq 6$ literature are $0.2-2~{\rm kpc}$, usually from ground-based campaigns with complementary HST or JWST imaging \citep[][]{Shibuya2014, Khusanova2020, Lemaux2021, Claeyssens2022, Ning2024}. To be physically interpreted, the spatial offsets should be correlated to the UV size of the galaxy \citep[e.g.,][]{Claeyssens2022}, avoiding possible biases that may cause bigger offsets to appear in larger galaxies. 

To achieve that aim, we normalize the offsets to the 90\%-light radius of the UV. {\ASL Since $\Delta_{\rm \lya - UV}/r_{90}^{\rm UV} \leq 1$ in all cases,} the centroids of the \lya\ emission in LaCOS appear to always be confined within the UV contours of the galaxy. {\asl The spatial coincidence of the UV and \lya\ centroids suggests that either the bulk of the \lya\ is primarily produced within the location star-forming regions, or that the diffuse \lya\ emission manifests symmetrically extended around them, compatible with the scattering scenario.} This is also consistent with cosmological simulations \citep[e.g.,][]{Lake2015, Byrohl2021, Mitchell2021}, that only predict important contributions from the other mechanisms (cooling, recombination or galaxy clustering) at distances in the halo well above the ones detected in individual LaCOS galaxies \citep[i.e., $\geq 10~{}\rm kpc$, see][]{LujanNiemeyer2022b, Guo2024}. Reassuringly, {\ASL and although a large fraction of the LaCOS galaxies show signatures of interactions or mergers near coalescence \citep[$\geq50\%$][]{LeReste2025-mergers}($\geq50\%$), we do not find clear evidence of separate companions emitting in \lya,} once again ruling out the galaxy clustering scenario. 

Turning to the literature, the results presented in \citet{Leclercq2024} concerning a subsample of LzLCS galaxies --which includes some of our LCEs and non-LCEs-- are compatible with this work, reporting offsets between the MgII and the stellar emission that did not extend beyond the size of the HST counterpart. Sharing a similar resonant nature, this confirms the ability of MgII to trace the same neutral and low-ionzed gas as \lya\ \citep[e.g.,][]{Henry2018, Chisholm2020, Xu2022}. Contrarily, around half of the $z=3-5$ lensed LAEs from \citet{Claeyssens2022} show much higher offsets than the UV size, which was attributed to the presence of {\ASL companions in their sample.}

\begin{figure}
    \centering
    \includegraphics[width=0.99\columnwidth, page=1]{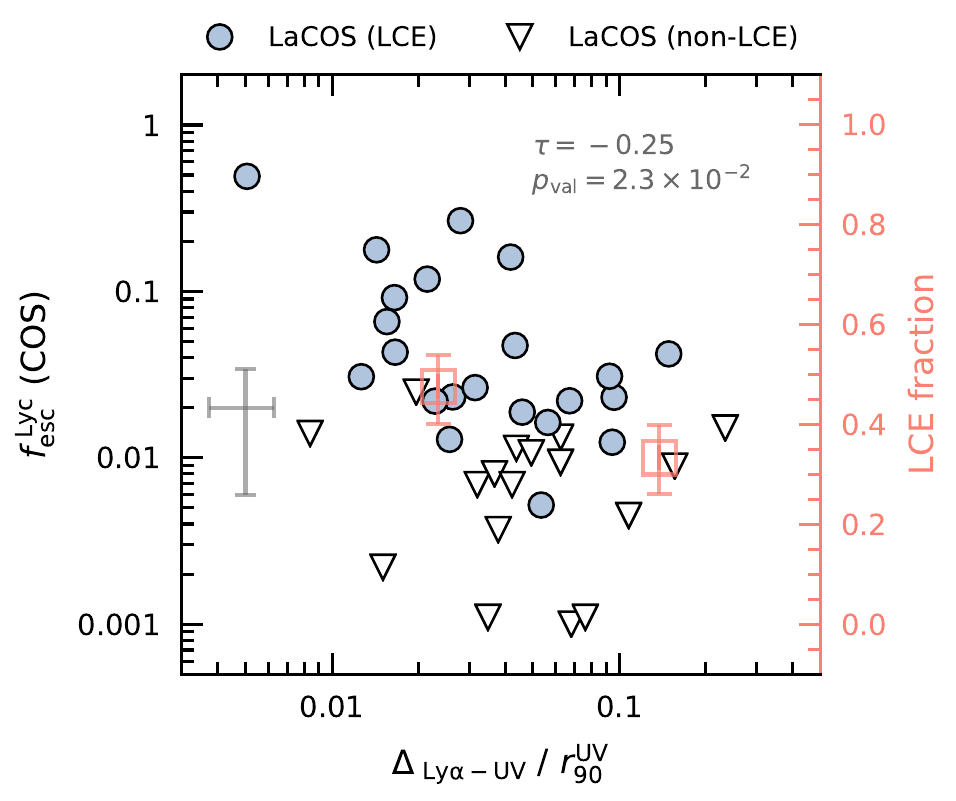}
\caption{{\bf Ionizing escape fraction (\fesclyc) as a function of the spatial offset between the centroid of the \lya\ and UV emission ($\Delta_{\rm \lya-UV}$), relative to the size of the UV continuum ($r_{90}$).} The centroid of the \lya\ appears confined within the UV emission for all LaCOS galaxies. LCEs show smaller relative offsets than non-LCEs (marginally), suggesting that both \lya\ and LyC preferentially escape through privileged sight-lines aligned with the observer.}
\label{fig:fesc_LyaUVoffset}
\end{figure}

In Figure \ref{fig:fesc_LyaUVoffset}, we show the escape fraction versus the \lya-UV offset relative to $r_{90}^{\rm UV}$, $\Delta_{\rm \lya - UV}/r_{90}^{\rm UV}$. Our observations reveal a tentative correlation, indicating that smaller relative offsets are found in galaxies with high \fesclyc, {\ASL a result that was already found (tentatively, albeit with a different method), in \citet{LeReste2025}. {\asl Similarly, \citet{Kim2023} observed a spatial coincidence of both \lya\ and LyC photon escape from a single star cluster in the Sunburst Arc at $z = 2.4$.} In the same vein, four out of the five strong LyC emitters ($\fesclyc \geq 20\%$) reported in \citet{Kerutt2024}, using HST/F336W photometry cross-matched with VLT/MUSE spectroscopy \citep{Inami2017}, showed spatial offsets almost coincident between \lya, UV and the LyC \citep[see also][]{MarquesChaves2024}. Other high-$z$ studies, on the other hand, has shown significant offsets of the LyC respect to the UV \citep{Fletcher2019, Ji2020}.} Finally, and motivated by the former works, \citet{Choustikov2024} studied the relation between \lya\ offsets and LyC escape in the SPHINX cosmological simulations \citep{Rosdahl2022}, and found that galaxies that contribute most to reionization tend to have $\Delta_{\rm \lya-UV} \leq 1{\rm kpc}$, although there was no clear trend between \fesclyc\ and $\Delta_{\rm \lya-UV}$.

\section{{\aref Properties of \lya\ halos (LAHs) around low-z, compact, star-forming galaxies}}\label{sec:results_modeling}
In the previous section, we have unveiled the presence of \lya\ emission with half-light radii three times larger than the corresponding size in the UV continuum on average (and up to six times in some cases; Fig.\,\ref{fig:LyaUV_size}). Based on the small offsets between the \lya\ and UV centroids and the lack of close galaxy companions, we argued that these \lya\ halos (LAHs) most likely originate from the scattering of \lya\ photons emitted from the star-forming regions into the extended HI halo of these galaxies \citep[e.g.][]{Steidel2011}\footnote{Even though, the possibility of a diffuse population of unresolved HII regions extending beyond the UV emitting contours cannot be discarded \citep[e.g,][]{OeyClarke1998}. Disentangling between both \lya\ production mechanisms will require follow-up observations of hydrogen recombination emission lines (e.g., H$\beta$, H$\alpha$).}. {\aref Furthermore, tentative differences in the morphology of the extended \lya\ emission have been found between LCE and non-LCE populations, where LCEs exhibit more concentrated \lya\ distributions than non-LCEs (Fig.\,\ref{fig:fesc_CLya}). However, simple size measurements could not accurately reproduce the morphology of \lya\ in these compact galaxies.} {\ASL To further test the underlying hypothesis of whether LCEs show more compact \lya\ respect to the UV than non-LCEs,} we proceed to model the shape, extension and luminosity of the LAHs in LaCOS by employing fitting methods widely used in  the literature. 

\begin{figure*}
    \centering
    \includegraphics[height=6cm, page=1]{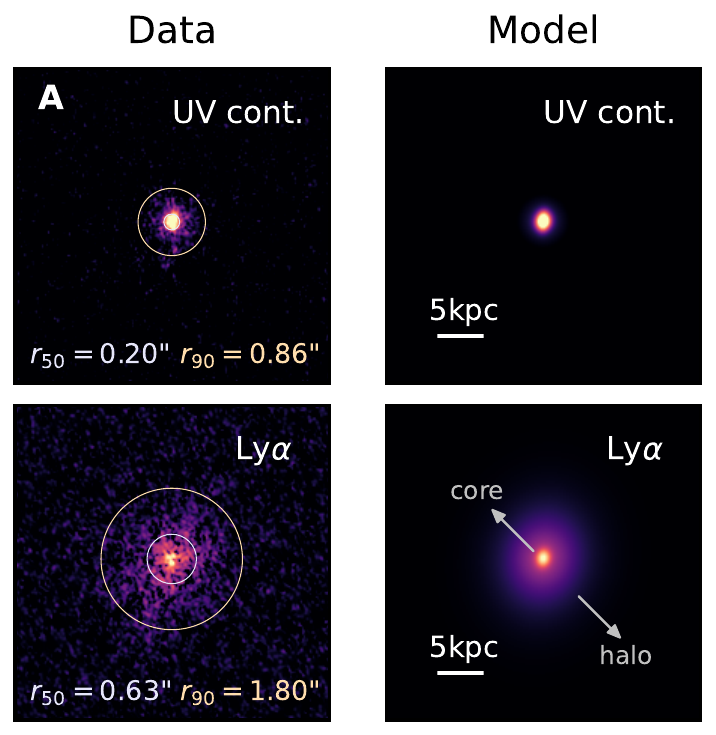}
    \includegraphics[height=6cm, page=1]{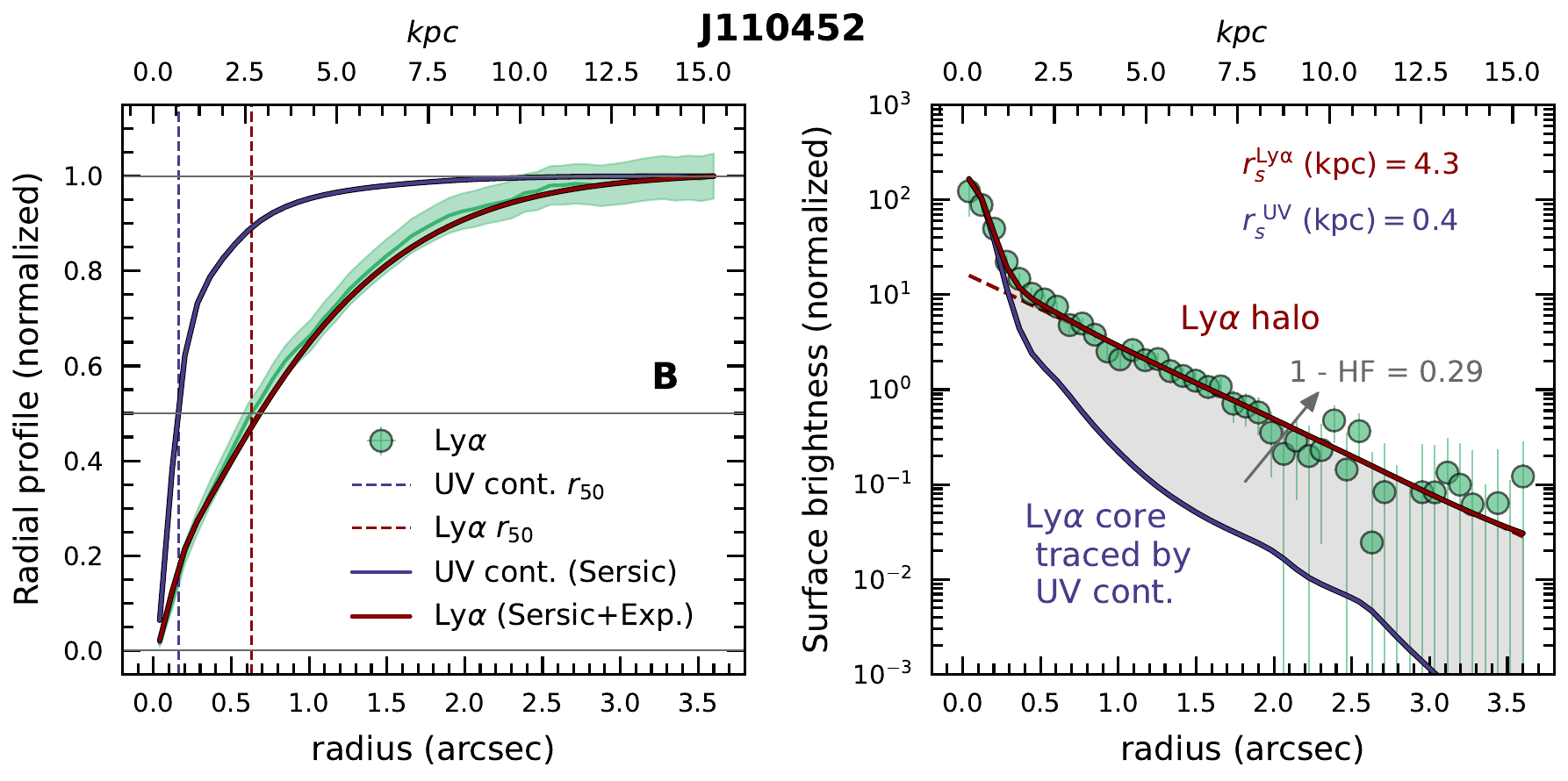}
\caption{{\bf Example of \lya\ halo modeling for the non-LCE galaxy J110452.} Panel (A): UV continuum and \lya\ images (left column), and corresponding best-fit \textsc{pysersic} models (right column). Concentric circles in blue and yellow mark the measured 50$\%$-light and 90$\%$-light radius on each band. The white bar corresponds to 5 physical kpc at the redshift of the source. Panel (B): \lya\ radial and surface brightness profiles (green shaded area and data points). The projected single Sersic and Sersic+Exponential models that fit the core (traced by the UV continuum) and {\ASL the core+halo emission of the \lya,} are shown with blue and red solid lines, respectively. The resulting UV and \lya\ scale lengths ($\rsUV, \rsLya$) can be read in the insets. The \lya\ Halo Fraction (HF), representing the integral of the halo component over the total \lya\ luminosity, is also shown.}
\label{fig:LAH_showcase}
\end{figure*}

\subsection{Modeling of LAHs in compact galaxies}
Extended LAHs have been shown to be ubiquitous around star-forming galaxies at all redshifts, detected via stacking techniques \citep[e.g.,][]{Hayashino2004, Steidel2011, Matsuda2012, Momose2014, Momose2016, Xue2017, Wisotzki2018, Kakuma2021, LujanNiemeyer2022a, LujanNiemeyer2022b, Kikuchihara2022, Kikuta2023, Guo2024, Zhang2024} and around individual galaxies \citep[e.g.,][]{Fynbo2001, Swinbank2007, Rauch2008, Hayes2013, Hayes2014, Patricio2016, Wisotzki2016, Leclercq2017, Erb2018, Kusakabe2019, Claeyssens2022, Kusakabe2022, Rasekh2022, Erb2023, Runnholm2023, Song2024}.

The 2D light distribution of LAHs have often been modeled assuming two morphological components \citep{Wisotzki2016, Leclercq2017, Rasekh2022, Runnholm2023}. The first component (named \emph{core}), steeper and more compact, traces the \lya\ photons directly produced within the central star clusters, and is assumed to match the shape of the UV counterpart. The second component (the \emph{halo}), often flatter and more extended than the core, probes the \lya\ within the CGM, whose morphology is independent of the shape of the core. Inspired by former studies \cite[e.g.,][]{Leclercq2017}, here we adopt the same two component fitting approach. 

We fit the spatial distribution of our synthetic NB \lya\ images using a 2D S\'ersic+Exponential profile decomposition of the form:
\begin{equation}
\begin{split}
I(x,y) \propto I_{\rm core}^0 \cdot \exp \left( \left( -\dfrac{r_{\rm core}(x_0,y_0,\theta,q)}{r_s^{\rm core}} \right)^{1/n} \right) + \\
+ I_{\rm halo}^0 \cdot \exp \left( -\dfrac{r_{\rm halo}(x_0,y_0)}{r_s^{\rm halo}} \right)
\end{split}
\end{equation}
\label{eq:LAH_model}

\noindent where $r_{\rm core}(x_0,y_0)$ is a rotated ellipse centered at $(x_0,y_0)$ with position angle $\theta \in [0,2\pi)$ (measured in radians from the positive $x-$axis), and axis ratio $q=1-b/a \in [0,1)$ (with $b, a$ the semi-major and semi-minor axes). $r_{\rm halo}(x_0,y_0) = \sqrt{(x-x_0)^2 + (y-y_0)^2}$, $x,y$ being the cartesian coordinates in pixel units. $r_s^{\rm core}, r_s^{\rm halo}$ are the characteristic core and halo scale lengths (in pixels), and $I_{\rm core}^0, I_{\rm halo}^0$ are the central intensities of the S\'ersic and Exponential profiles (in flux density units).  

The fits are performed using the the Bayesian code \textsc{pysersic} \citep{pysersic}, which allows for a flexible control of the priors while taking into account the instrumental PSF by convolving the models with our custom PSF kernel (Sect.\,\ref{sec:data}). During the fit, we enforce $r_s^{\rm core} \leq r_s^{\rm halo}$, while $r_s^{\rm core}, n, \theta, q$ {\asl and the ellipse centroid $(x_0,y_0)$} are fixed to the best solution obtained from a separate S\'ersic fit to the UV continuum image alone. $I_{\rm core}^0$ is free to vary so that the intensity of the core scales to the luminosity of the central \lya\ component. Figure \ref{fig:LAH_showcase} shows an example of our LAH modeling approach. Panel (A) shows the data and best-fit 2D models for the UV continuum (top) and NB \lya\ (bottom). In the case of \lya, the core and halo components have been highlighted. In Panel (B), the best-fit models for the UV (in blue) and \lya\ (in red) are projected into circularized radial and surface brightness profiles, together with the observed \lya\ distributions (in green). {\ASL The need for the extended halo component to capture the light of the outer regions of the \lya\ emission is clear.}

In Table \ref{tab:LAH_analysis}, we report the mean and inter-quartile range of the \textsc{pysersic} realizations for the core (UV) and \lya\ halo scale lengths. We obtain $r_s^{\rm core} \equiv r_s^{\rm UV} = 0.11-1.97{\rm ~kpc}$ and $r_s^{\rm halo} \equiv r_s^{\rm \lya} = 0.93-7.61{\rm ~kpc}$. Figure \ref{fig:LyaUV_scale} shows the comparison between the \lya\ and UV scale lengths for LaCOS galaxies. {\asl LaCOS LAHs extend, overall, 10 times beyond the size of the UV starlight, effectively probing distances out to the CGM of these compact galaxies.} As a comparison, in Fig.\,\ref{fig:LyaUV_scale} we include the LAH measurements from the MUSE-Deep survey at $z = 3-5$ \citep{Leclercq2017}, and the results from the eLARS nearby galaxy sample \citep{Rasekh2022}. 

{\aref Clearly, the extent of LaCOS LAHs is in agreement with the MUSE results at higher redshifts, with some of the eLARS galaxies being more extended systems in the UV. \citet{Runnholm2023} first interpreted these similarities as a potential lack of evolution in the relative sizes of extended \lya\ with cosmic time. Here, we additionally stress the high-$z$ analog nature of the LaCOS galaxies to explain this behavior. The high (and compact) SFRs, blue UV colors and extreme emission line properties of the LaCOS galaxies are properties shared among the high-$z$ galaxy population \citep[e.g., see][]{Mascia2024}. Therefore, similar properties between the LaCOS analogs and the MUSE high-$z$ LAEs will lead to similar LAHs, regardless of the redshift, given the underlying connection between the latter and various galaxy properties \citep[e.g.,][]{Rasekh2022}.}

\begin{figure}
    \centering
    \includegraphics[width=0.99\columnwidth, page=1]{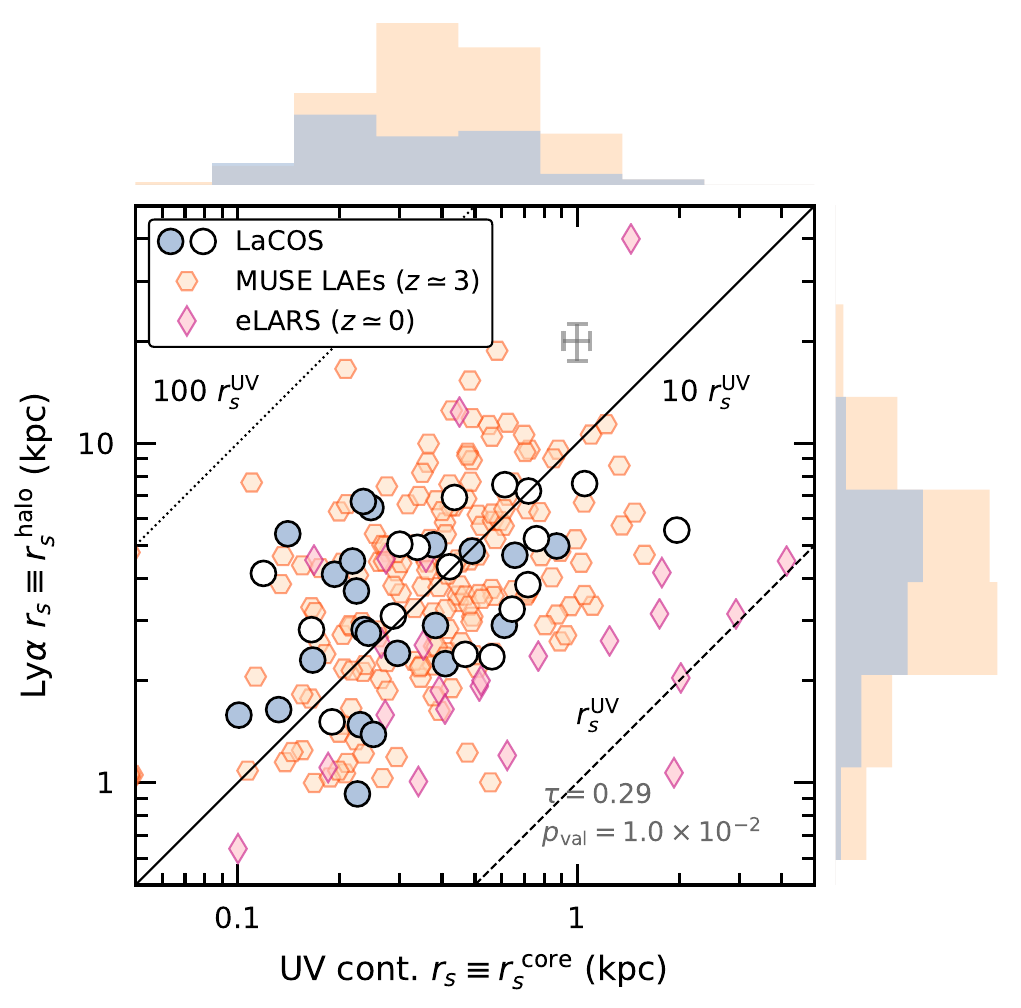}
\caption{{\bf \lya\ halo versus UV continuum scale lengths.} {\asl Solid and empty circles show LaCOS LyC detections and upper limits, respectively.} Results from the MUSE \citep{Leclercq2017} and eLARS surveys \citep{Rasekh2022} are shown via orange and pink symbols. The dotted, solid and dashed lines draw the $\rsUV, 10~\rsUV$ and $100~\rsUV$ equalities, respectively. Both nearby star-forming galaxies and high-$z$ LAEs show \lya\ halos that are around ten times larger than the characteristic UV emission, hinting on the lack of evolution in the distribution of neutral CGM gas with cosmic time.}
\label{fig:LyaUV_scale}
\end{figure}

\subsection{The \lya\ Halo Fraction (HF)}
Another commonly used quantity to characterize LAHs is the so-called \lya\ halo fraction (HF). This parameter represents the contribution of the halo to the total \lya\ luminosity \citep[e.g.,][]{Steidel2011, Wisotzki2016, Leclercq2017}, and it is defined as:
\begin{equation}
    {\rm HF} = \dfrac{L_{\rm halo}^{\rm \lya}}{L_{\rm halo}^{\rm \lya}+L_{\rm core}^{\rm \lya}}
\end{equation}

\noindent where $L_{\rm core}^{\rm \lya}$ and $L_{\rm halo}^{\rm \lya}$ correspond to the integrated luminosity of the core and the halo, respectively. A compilation of our HFs for the LaCOS galaxies can be found in Table \ref{tab:LAH_analysis}. We obtain HFs ranging from 0.1 for the faintest and more compact halos, to 0.9 for the more luminous and extended ones, compatible with MUSE results. The behavior of the HF with other halo-related quantities has been widely studied in \citet{Wisotzki2016} and \citet{Leclercq2017}. For example, while the halo luminosity ($L_{\rm halo}^{\rm \lya}$) seems to scale with the UV and \lya\ scale lengths, the HF does not appear to correlate with these quantities. Once again, our results agree with the former studies. {\aref The lack of discernible differences between our $z \simeq 0.3$ LAHs and the MUSE measurements at $z \geq 3$ implies that the physical conditions of the neutral CGM is similar between these high-$z$ galaxies and our analog sample.} This supports the applicability of \fesclyc\ predictors that rely on the spatial properties of \lya\ to observations of high-$z$ galaxies. In the same line, \citet{Roy2023} recently reported similar physical and \lya\ characteristics between the well-studied local analogs and a sample of 11 high-$z$ LAEs with combined JWST plus MUSE observations.

\begin{figure}
    \centering
    \includegraphics[width=0.99\columnwidth, page=1]{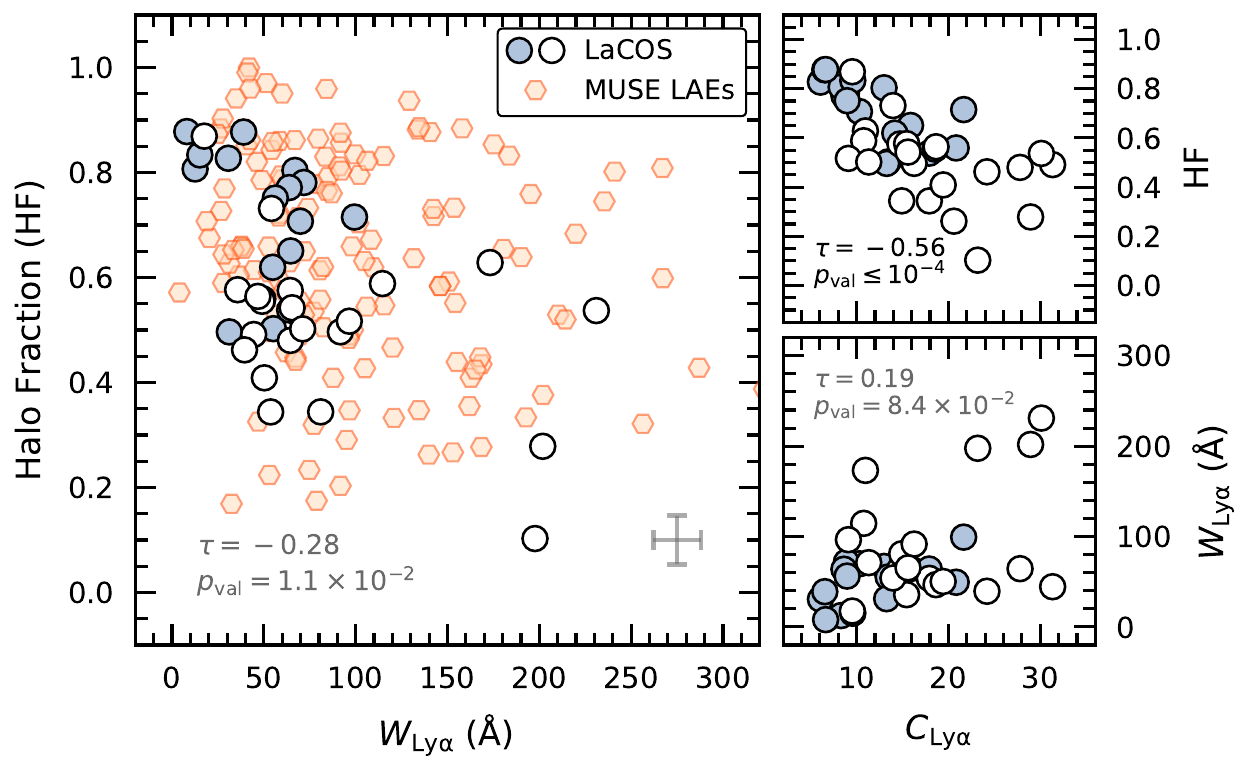}
\caption{\aref {\bf \lya\ Halo Fraction (HF) as a function of the \lya\ equivalent width and the \lya\ concentration parameter ($C_{\lya}$).} Legend is an in Fig.\,\ref{fig:LyaUV_scale}. The HF and \ewlya\ are mildly anti-correlated, while the HF decreases sharply with $C_{\lya}$. These trends imply that the \lya\ flux in the strongest LAEs emerges, mainly, from the central starbursts rather than from the diffuse halos.}
\label{fig:HF_LyaLyC}
\end{figure}

For the topic of this work, it is interesting to explicitly show how these HFs behave with global integrated properties of the \lya\ line \citep[e.g.,][]{Leclercq2020}. As summarized in the Introduction of this paper, both \ewlya\ and \fesclya\ hold some of the strongest scaling relations with \fesclyc \citep[e.g.,][]{Izotov2020, Pahl2021, Flury2022b}. Strong LAEs (i.e., high \ewlya) will be, statistically speaking, strong LyC emitters too \citep[e.g.,][]{Steidel2018, Izotov2024}. {\aref Figure \ref{fig:HF_LyaLyC} shows the HF against the total equivalent width of \lya\ as well as against the concentration parameter ($C_{\lya}$) described in Sect.\,\ref{sec:results_nonparam} (for completeness, we also plot the latter two quantities against each other).} For comparison, the high-$z$ LAH measurements from MUSE \citep{Leclercq2017} are shown in the background of this plot. Contrarily to the high-$z$ observations, LAHs in LaCOS do show a marginal anti-correlation between the HF and \ewlya. {\asl On the other hand, HF appears to significantly scale with $C_{\lya}$ ($p_{\rm val.} \leq 10^{-4}$), showing that low HF corresponds to highly concentrated halos (high $C_{\lya}$), and vice versa.} This behavior suggest that most of the \lya\ flux contributing to the \lya\ equivalent width in LAEs actually originates from the central starburst (high $L_{\rm core}^{\rm \lya}$) rather than from the diffuse emission in the halo \citep[low $L_{\rm halo}^{\rm \lya}$, see][]{Steidel2011, Wisotzki2016}. {\asl Consistent with this interpretation, \citet{Blaizot2023} found that the inner CGM acts as a screen that scatters some \lya\ photons out of the line of sight, producing a effective absorption in the line profile. As a consequence, larger HF may imply a more extended CGM in front of the galaxy which is detectable in \lya\ emission, but the overall \lya\ flux is strongly reduced (low EW), whereas when the LAH is compact, most of the produced \lya\ flux can be transmitted.}

Finally, it is worth noticing the different space of parameters occupied by LCEs and non-LCEs in Fig.\,\ref{fig:HF_LyaLyC}. {\asl LCEs seem to have {\ASL lower HFs than non-LCEs} which points towards a situation in which the majority of both \lya\ and LyC would escape either (1) straight from the star clusters and through privileged sight-lines towards the observer, or (2) isotopically in all directions.} {\ASL Consistently with the first interpretation, in \citet{LeReste2025} we found a strong degree of correlation between \fesclyc\ and the \lya\ luminosity and equivalent width of the brightest UV-emitting clusters in LaCOS, suggesting that the escaping LyC radiation preferentially originates from the brightest clusters in the galaxies, and further supporting the connection between \lya\ properties and LyC escape.} 

Armed with the HF as our primary metric to characterize the \lya\ halos, in the next section we will address the fundamental question aim by the LaCOS program: \emph{how do conditions in the CGM, as traced by \lya, impact LyC escape in galaxies?} 

\section{{\aref On the connection between LyC escape and the properties of LAHs in emission}}\label{sec:discussion}
The main goal of this paper is to establish the connection between the properties of the extended CGM (probed by \lya\ emission) and the physics of LyC escape. To do so, throughout we have characterized the LAHs in a sample of 42 galaxies with ionizing continuum observations: the LaCOS sample. In the previous section, we defined the Halo Fraction (HF) as the fractional contribution of the halo to the total \lya\ luminosity. Now, we study the relation between HF, \fesclyc\ and the physical properties of LaCOS galaxies. 

\begin{figure*}
    \centering
    \includegraphics[width=0.80\textwidth, page=1]{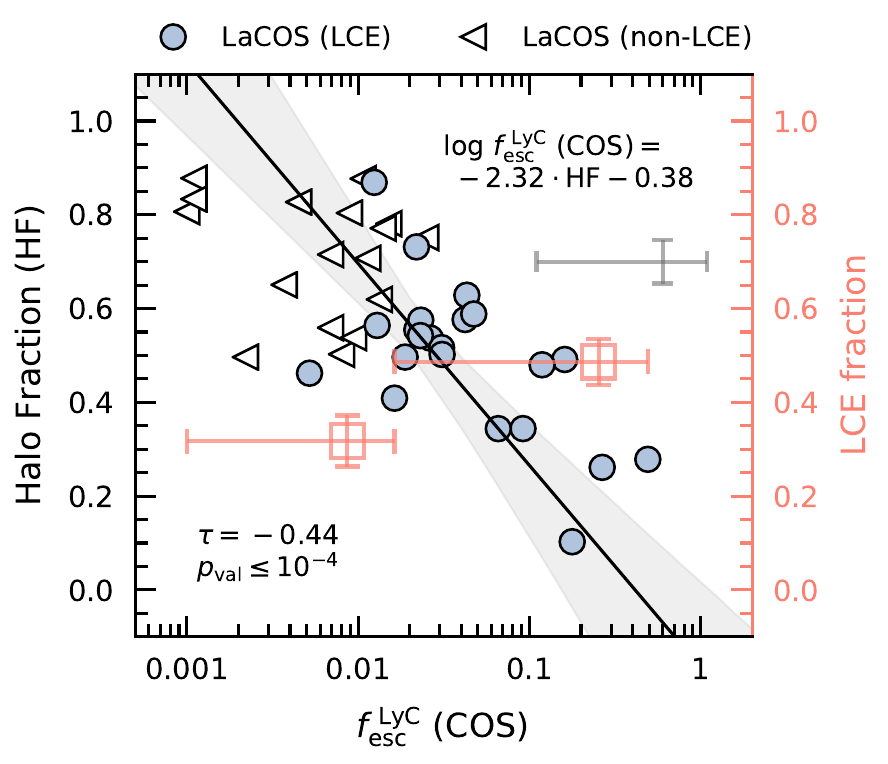}
\caption{{\bf Relation between the \lya\ Halo Fraction (HF) and the ionizing escape fraction (\fesclyc) in the LaCOS sample.} The solid line represents a linear fit to the data, including upper limits. LCEs and galaxies with high \fesclyc\ show lower HFs than non-LCEs, indicating that \lya\ and LyC escape from the central star clusters and, in the case of \lya\ radiation, {\ASL minimizing the number of scattering interactions in the intervening CGM.}}
\label{fig:HF_fesc}
\end{figure*}

\subsection{The HF to \fesclyc\ relation}
Figure \ref{fig:HF_fesc} shows the relation between the \lya\ Halo Fraction (HF) and the ionizing escape fraction (\fesclyc) in the LaCOS sample. Our Kendall ranking test reveals a strong ($\tau=-0.44$) and significant ($p_{\rm val.} \leq 10^{-4}$) anti-correlation between the two, so that galaxies with high \fesclyc\ show low HFs, and vice-versa. Our findings imply that both the \lya\ and the ionizing radiation in LCEs emerge directly from the central star-forming regions (high $L_{\rm core}$), a physical picture which is consistent with the already found correlations between \fesclyc\ and the \lya\ properties of the UV-brightest clusters in LaCOS \citep{LeReste2025}, {\asl as well as other LCEs at higher redshifts \citep{Kim2023}.} As this \lya\ halo is most likely produced via scattering within the CGM gas in the line of sight (see previous sections), \lya\ photons in LCEs (high \fesclyc) would escape {\ASL without much scattering interactions} in the surrounding CGM (lower $L_{\rm halo}$ and HFs). 

Based on radiative transfer simulations, \citet{MasRibas2017} first suggested that extended \lya, H$\alpha$ and UV continuum emission can be used to infer the escape fraction of ionizing radiation from a central source into the CGM. \citet{Choustikov2024} built on this, and used mock observations from the SPHINX cosmological simulations \citep{sphinx, Rosdahl2022, Katz2023} to show that galaxies with larger angle-averaged \fesclyc\ tend to have less extended \lya\ profiles with respect to both the rest-UV continuum and H$\alpha$ emissions. {\ASL However, changes in UV extent were smaller than those for H$\alpha$ with respect to \lya, probably due to fluorescence exciting H$\alpha$ emission in the outer CGM, while the UV profiles become increasingly steep due to the presence of nuclear starbursts.} Our results support the aforementioned simulations, and the tight correlation found between HF and the escape fraction motivate the use, for the first time, of the HF as a new \fesclyc\ indicator. 

\begin{figure*}
    \centering
    \includegraphics[width=0.95\textwidth, page=1]{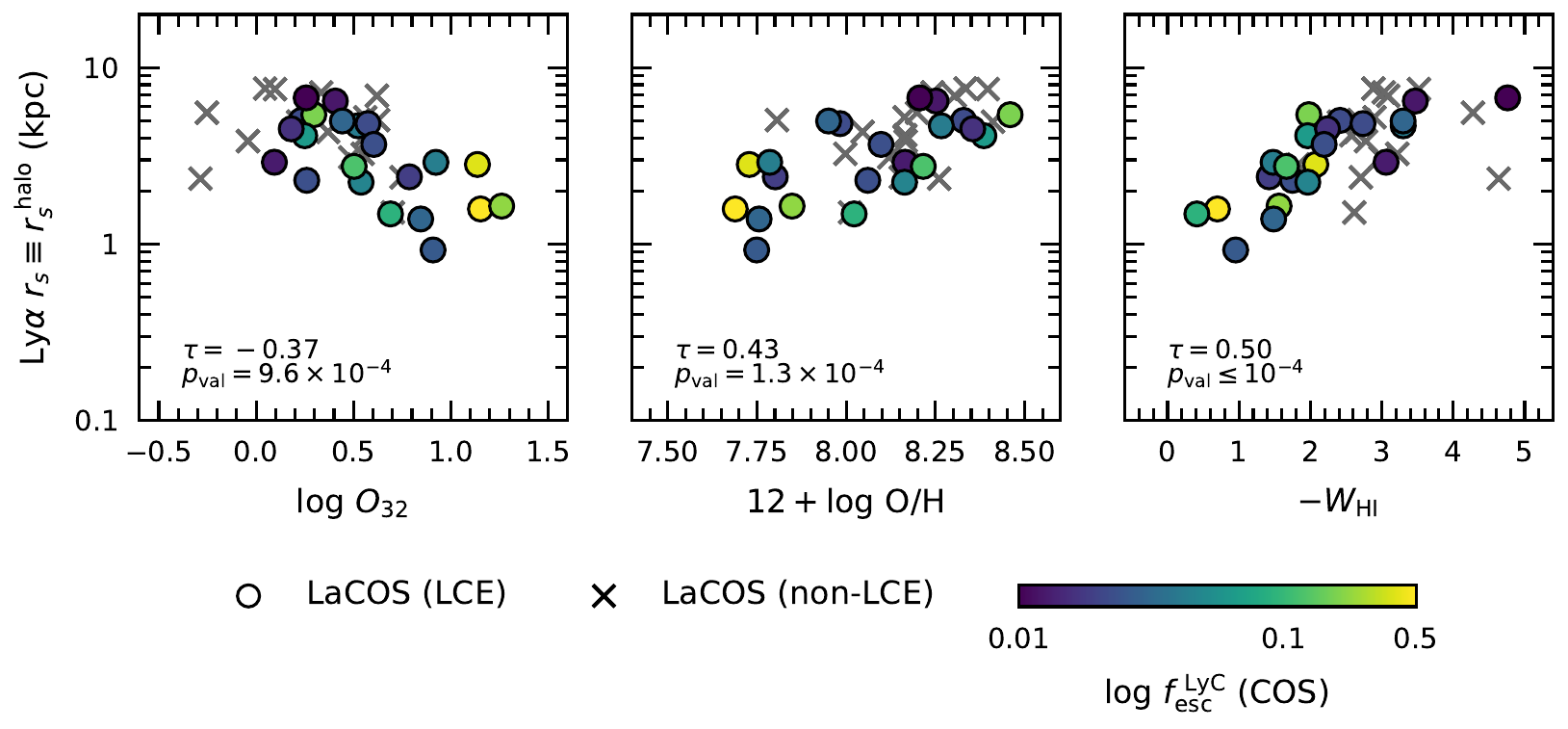}
\caption{{\bf The dependence of the \lya\ scale length (\rsLya) on the LaCOS physical properties:} the ionization parameter (traced by $\log O_{32}$), the gas-phase metallically ($12 + \log {\rm O/H}$), and the equivalent width of the HI lines ($\ewhi$, a proxy for the line-of-sight HI column density). Data points are color-coded by \fesclyc, {\asl and crosses represent non-detections in the LyC.}}
\label{fig:rsLya_props}
\end{figure*}

We fit a linear regression model {\asl (the simplest we found that could accurately describe the empirical trend)} to the $\log \fesclyc$ versus HF observations using \textsc{linmix} \citep{linmix}, including errors on both variables and accounting for censored data. We obtain:
\begin{equation}
	\log \fesclyc = (-2.32 \pm 0.41) \cdot {\rm HF} - (0.38 \pm 0.25)
\end{equation}

\noindent with a resulting small intrinsic scatter of $\sigma_y = 0.02$. Albeit the non-negligible uncertainties, HF can be used to estimate the escape fraction of {\ASL \lya-emitting galaxies at high-$z$. 

{\aref As an illustrative example, we use four of the $z \geq 3$ LyC detections discovered by \citet{Kerutt2024} using the HDUV survey \citep[][including the UVUDF region]{Oesch2018}, with IDs 7193, 1087 (Gold), 2134 and 7121 (Silver sample), and available MUSE-Deep data. Our predicted \fesclyc\ from the estimated HFs in \citet{Leclercq2017} is only 1-2\% (${\rm HF}\simeq 0.6-0.8$), while other \lya\ observables such as high \lya\ peak separations (677 and 565 ${\rm ~km ~s^{-1}}$ in the case of IDs 1087 and 2134), suggest negligible escape too. Puzzlingly, \citet{Kerutt2024} reported \fesclyc\ values between 20 and 80\% for these sources (with $\pm 5-15\%$ typical uncertainties). The discrepancy may arise from plausible under-estimations of the intrinsic LyC flux from the broad-band SED modeling of the HDUV sources.} 

This is specially relevant at the EoR, where the only accessible \fesclyc\ information needs to imperatively arrive from indirect diagnostics \citep[e.g.,][]{Jaskot2024b}. Luckily, the number of \lya\ observations within the EoR is growing at unprecedented pace thanks to JWST \citep[e.g.,][]{Bunker2023, Roy2023, Saxena2023, Tang2023, Jones2024, Jung2024, Napolitano2024, Saxena2024, Tang2024, Witstok2024, Witten2024, Jones2025, Runnholm2025, Witstok2025}. However, the former works are based on integrated spectral measurements at ISM scales, {\ASL which are impacted by the IGM absorption at these epochs. Relative morphological properties of the \lya, such as HFs, will presumably be less affected by the IGM, given the difference in physical scales between these features and the ionized bubbles at the EoR \citep[e.g.,][]{HayesScarlata2023, Lu2024}.} {\asl If the halo emission is more redshifted than the core emission due to outflows, for example, this may hamper the direct use of the LAH diagnostics, since the halo emission will be less attenuated by the IGM than the \lya\ photon from the core \citep[see e.g.,][]{Garel2021}.}

As a forecast for future studies, NIRSpec/IFU observations will be able to detect and characterize the extended LAHs around these distant sources, with a resolution below $600{\rm ~pc}$ at $z \simeq 6$. This opens a new window for the reionization community so that, by using the relation between \fesclyc\ and HF proposed in this work, the ionizing output of galaxies can be estimated at the EoR and beyond (Saldana-Lopez et al., in prep.). {\asl Even for high-$z$ LAEs with no detected UV counterpart, the strong correlation between the HF and the \lya\ concentration ($C_{\lya}$), will allow to alternatively use the latter as a proxy for the LyC escape fraction.}

\begin{figure*}
    \centering
    \includegraphics[width=0.95\textwidth, page=1]{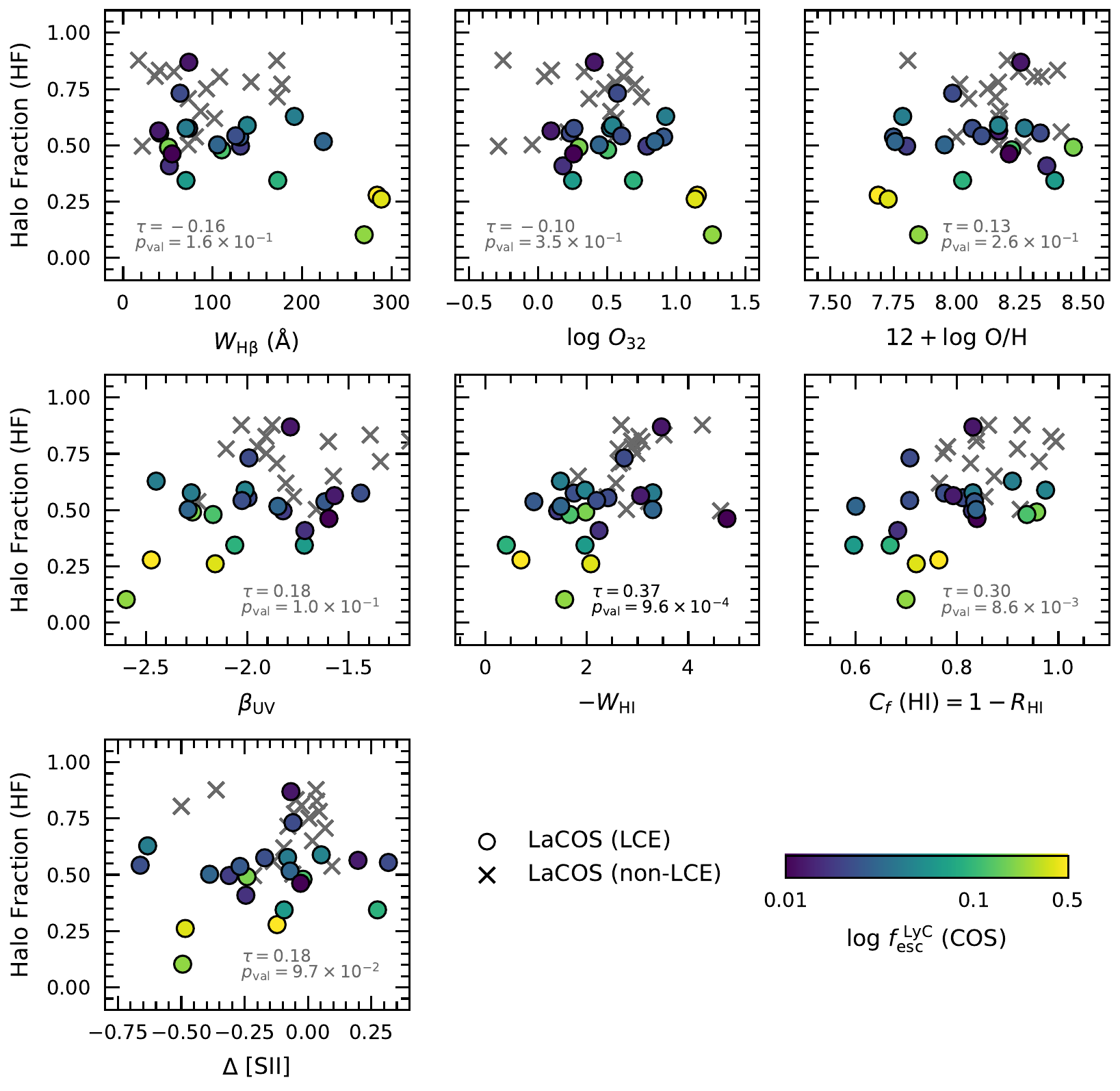}
\caption{{\bf \lya\ Halo Fraction (HF) versus physical properties of LaCOS galaxies}. HF is plotted against different parameters known to be indirect drivers of LyC escape, and tracing the youth of the stellar populations ($W_{\rm H\beta}$), ionization parameter ($O_{32}$), the metallicity of the gas ($12 + \log {\rm O/H}$), dust attenuation ($\beta_{\rm UV}$), the density and covering fraction of the HI gas ($W_{\rm HI}$, $R_{\rm HI}$), {\asl and matter-bounded versus radiation-bounded galaxies ($\Delta[{\rm SII]}$).} Data points are color-coded by \fesclyc, aiming to highlight the connection between HF, LyC escape and the physical properties of these nearby LAEs.}
\label{fig:HF_props}
\end{figure*}

\subsection{The role of the neutral CGM in LyC escape}
Here, we study the physical galaxy parameters that could impact both the extent and contribution of the halo to the total \lya\ luminosity. To do so, we compare the \lya\ scale lengths (\rsLya, Figure \ref{fig:rsLya_props}) and HFs in LaCOS (Figure \ref{fig:HF_props}) with some of the parameters known to be indirect drivers of LyC escape \citep[e.g.,][]{Izotov2021, Flury2022b, SaldanaLopez2022}. Specifically, we use the equivalent width of the Balmer lines (i.e., $W_{\rm H\beta}$) as an indicator of the age of the stellar populations, the $O_{32}$ ratio as a proxy for the ionization parameter, and $12+\log{\rm ~O/H}$ for the gas-phase metallicity. We will also employ the UV continuum slope ($\beta_{\rm UV}$) as a tracer of dust attenuation, and the equivalent width and residual flux of the Lyman series lines ($W_{\rm HI}, R_{\rm HI} = 1 - C_f{\rm (HI)}$) for the density and covering fraction of the ISM HI gas. {\asl Finally, we will use the deficit in the [SII]$\lambda\lambda$6716,6732/H$\alpha$ ratio ($\Delta[{\rm SII}]$) respect to the bulk of SDSS star-forming galaxies \citep{Wang2019}, as a proxy for matter (or [SII]-deficient) versus ionization-bounded galaxies.}

In agreement with the results presented in \citet{Leclercq2024} studying MgII halos, Fig.\,\ref{fig:rsLya_props} shows that compact \lya\ configurations are preferentially found in galaxies with high ionization parameter, low metallicity and low HI equivalent width {\asl (measured from COS in the Lyman series)}. This suggests that either (1) the stellar populations in these galaxies have efficiently ionized not only the ISM but also part of their neutral gas halo \citep[e.g.,][]{Flury2025}, {\ASL and/or (2) \lya\ escapes through many sight-lines in all directions, consistent with the high LyC detection fraction in high $O_{32}$ galaxies \citep[e.g.,][]{Izotov2018b, Kim2023}.} These arguments lie in agreement with the findings by \citet{Kanekar2021} and \citet{Chandola2024}, who reported a low HI 21cm detection rate in nearby, low-mass galaxies with high $O_{32}$ ratios \citep[e.g., see also][]{McKinney2019}. We note, however, that in the case of single-dish studies, the low angular resolution of observations may play a role in the low detection rate of these compact galaxies. 

{\asl In any case, the correlation between the \lya\ scale length (\rsLya) and the HI equivalent width (\ewhi), points to a direct link between the \lya\ extended emission and the HI gas in front of the UV-emitting regions.} Continuing, we find a lack of correlation between \rsLya\ and the dust attenuation ($\beta_{\rm UV}$), in agreement with other observational studies \citep{Rasekh2022}. We caution that these correlations can be driven, at least to some extent, by the fact that galaxies with larger UV counterparts may have higher \lya\ scale lengths as well (Fig.\,\ref{fig:LyaUV_scale}). This way, disentangling the role of the CGM from other underlying scaling relations may be a difficult task. 

Fortunately, \lya\ HFs are independent of the UV or \lya\ scale lengths \citep{Leclercq2017, Rasekh2022}, while still being a good representation of the contribution of the halo. Fig.\,\ref{fig:HF_props} depicts a lack of correlation between the HF and properties related with the stellar populations ($W_{\rm H\beta}, O_{32}$), the dust ($\beta_{UV}$), {\asl or the [SII] deficit}. {\asl It is worth noticing, however, that the three most extreme LCEs in our sample, having the highest $W_{\rm H\beta}, O_{32}$, the lowest ${\rm O/H}$ and $\beta_{\rm UV}$ and being among the most [SII] deficient, all show very low HFs.} {\asl Anyway, the lack of correlation between HFs and the above-mentioned physical quantities suggests that the overall properties of CGM are independent of those tracing the stellar populations at galactic scales}. On the other hand, we report significant correlations between the HFs and physical quantities related with the neutral ISM gas. In particular, higher \lya\ halo fractions are found for galaxies with higher $W_{\rm HI}$, although only a tentaive correlation is found with $C_f{\rm (HI)}$. {\asl This shows that properties that trace the optical depth of the HI gas within the ISM are linked to those of the LAHs seen in emission.} 

Based on stacked measurements of LzLCS spectra, \citet{Flury2025} found evidence for the concurrence of two LyC escape scenarios in galaxies. {\asl Conveniently, \citet{Carr2025} studied the effect of both radiation and mechanical feedback in LzLCS galaxies, by looking at the outflow profile of the absorption lines.} On the one hand, in the strongest leakers ($\fesclyc \geq 5\%$), stellar populations younger than 3~Myr increase the ionizing feedback, which in turn can foster the isotropy of LyC escape by fully ionizing the galaxy surroundings. {\asl Alternatively, in the lowest metallicity starbursts, the onset of SN can be further delayed or suppressed \citep{JecmenOey2023}, and catastrophic cooling may cause the build up of cool clouds at small radii, fragmenting into low density channels that the intense ionization front will rapidly evacuate.} On the other hand, in weak to moderate leakers ($\fesclyc < 5\%$), mechanical feedback from supernovae in 8-10~Myr stellar populations \citep{Bait2024}, imprint anisotropies in the dense gas distribution through which ionizing photons {\asl from subsequent starburst episodes can escape.} Crucially, the intensity of HI absorption lines in the spectra probe the density of the neutral gas in the ISM \citep[e.g.,][]{Gazagnes2018, Steidel2018} or, equivalently, the LyC optical depth of the ISM \citep[e.g.,][]{Flury2025}, so that galaxies with high \fesclyc\ also show weak HI lines \citep[][]{SaldanaLopez2022}. These HI gas indicators, together with \lya, have demonstrated to be one of the most promising proxies of LyC escape \citep[e.g.,][]{Jaskot2024a}. 

Our observations of LAHs in LyC emitting galaxies are consistent with the physical picture described in the paragraph above. Strong leakers host a highly ionized ISM with lower HI column densities, as evidenced by their high $O_{32}$ and low HI equivalent widths. {\asl The dominant ionizing feedback in the stronger LCEs struggles to drive gas into the CGM \citep[e.g.,][]{Jaskot2017}, resulting in more compact and shallow HI halos with shorter \lya\ scales and low HFs,} that would allow the LyC photons that escape the ISM to transfer through the CGM without being absorbed, while \lya\ photons would also evade significant scattering. In the weak and non-leaker regime, galaxies show high \lya\ scale lengths ($\rsLya \geq 2{\rm ~kpc}$) and high HFs (${\rm HF} \geq 0.5$), as well as high HI equivalent widths ($W_{\rm HI} \geq 2$\AA). This indicates a high HI column density of gas in front of the stars, {\asl probably as a result of intense galactic winds driving gas out to CGM scales \citep[e.g.,][]{Carr2025}.} The low fraction of \lya\ photons that do escape the ISM are likely to undergo significant scattering in the CGM producing a large \lya\ halo in emission, while the LyC radiation remain trapped before reaching the CGM and escape the galaxy. 

Apart from the outflows scenario \citep[see also][]{Zastrow2013, Amorin2024, Ferrara2025, Ji2025}, the confluence of both an optically thin ISM and a shallow neutral CGM in the line-of-sight, may alternatively be caused by either the stellar populations ionizing most of the neutral gas \citep[e.g.,][]{JaskotOey2013, Jaskot2019} or by the tidal forces of galaxy mergers \citep[e.g.,][]{Dutta2024, LeReste2024}.

\section{Summary and conclusions}\label{sec:conclusions}
In this paper, we have established the connection between the escape of ionizing radiation in galaxies and the physical conditions of the neutral gas in the the ISM and CGM. We have used data from the LaCOS program \citep{LeReste2025}, that provided \lya\ and UV continuum imaging (see Sect.\,\ref{sec:data}) for a sample of 42 low-redshift ($z \simeq 0.3$) star-forming galaxies with LyC observations \citep{Flury2022a}. Throughout, we have studied the size and morphology of the extended \lya\ emission compared to the UV counterpart (Sect.\,\ref{sec:results_nonparam}), and model the shape and contribution of \lya\ halos (LAH) to the total \lya\ luminosity (Sect.\,\ref{sec:results_modeling}). Finally, we have unveiled the relation between the LyC escape fraction (\fesclyc), the properties of LAHs, and the physical parameters that drive the escape of ionizing photons in LaCOS galaxies (Sect.\,\ref{sec:discussion}). The main conclusion of this article are summarized below. 

\begin{itemize}
    \item[$\circ$] LaCOS galaxies show extended \lya\ emission ubiquitously, with \lya\ half-light radius $\simeq 3$ times larger than the corresponding size of the UV continuum (Fig.\,\ref{fig:lacos_lahs_rgb} and \ref{fig:LyaUV_size}), on average, and \lya\ significantly detected at distances as far as 10 times from the UV starlight. This is in agreement with other studies of local star-forming galaxies \citep[e.g.,][]{Hayes2013, Melinder2023}. 
    \item[$\circ$] {\aref The reported anticorrelations between \fesclyc\ and both the \lya\ and UV size, seem to indicate that LCEs may have more compact \lya\ than non-LCEs respect to the UV continuum \citep[see][for a similar conclusion using MgII]{Leclercq2024}. However, individual data points do not reflect this behavior (Fig.\,\ref{fig:fesc_Lya}), highlighting the lack of ability of simple size measurements to reproduce the \lya\ morphology.} {\asl In any case, LCEs show more compact \lya\ light distributions than non-LCEs (Fig.\,\ref{fig:fesc_CLya}), where the centroid of the \lya\ is always confined within the UV contours (Fig.\,\ref{fig:fesc_LyaUVoffset}).}
    \item[$\circ$] The results of our 2D modeling and decomposition of the \lya\ emission in LaCOS (Fig.\,\ref{fig:LAH_showcase}), reveals LAHs with halo scale lengths that are $\simeq 10$ times more extended that the star-forming regions in the core (Fig.\,\ref{fig:LyaUV_scale}). {\aref These facts lay in agreement with measurements of LAHs at higher redshifts \citep[e.g.,][]{Leclercq2017}, and reinforces the idea of the LaCOS galaxies being robust analogs of high-$z$ emission line galaxy samples \citep[see also][]{Runnholm2023, Mascia2024}.}
    \item[$\circ$] We use the \lya\ halo fraction (HF) as the primary metric to characterize LAHs in LaCOS (Fig.\,\ref{fig:HF_LyaLyC}). These HFs, defined as the contribution of the halos to the total \lya\ luminosities, seem to be marginally lower for galaxies with high \ewlya\ (i.e., strong \lya\ emitters, LAEs), {\asl while they scale inversely with the \lya\ concentration ($C_{\lya}$)}. This suggests that the bulk of the \lya\ flux in strong LAEs mainly emerges from the central star clusters rather than from the diffuse outskirts of the halo \citep[e.g.,][]{Steidel2011, Wisotzki2016}. 
    \item[$\circ$] We discover an anti-correlation between the \lya\ HF and the escape fraction of ionizing photons (\fesclyc), so that LCEs and galaxies with high \fesclyc\ also have low HFs (Fig.\,\ref{fig:HF_fesc}). With this, we corroborate the results by \citet{Choustikov2024} based on cosmological simulations, and we propose the study of LAHs and the HF as new LyC escape indicators. {\ASL The resemblance between LaCOS and other high-$z$ surveys in the properties of LAHs, supports the applicability of these indicators to observations of high-redshift galaxies \citep[e.g.,][]{Roy2023}.}
    \item[$\circ$] Finally, we investigate other physical properties that may lead to the connection between LAHs and LyC escape (Fig.\,\ref{fig:rsLya_props} and \ref{fig:HF_props}). Specifically, the \lya\ scale length appear to decrease with the ionization parameter (traced by $O_{32}$), while it increases with the galaxy gas-phase metallicity ($12 + \log {\rm ~O/H}$). Furthermore, we report significant correlations between \rsLya, HFs and physical quantities related with the neutral gas in the ISM, so that higher \lya\ scale lengths and HFs are found for galaxies with higher HI equivalent widths of the Lyman series ($W_{\rm HI}$). 
    \item[$\circ$] In synthesis, we propose a physical scenario in which both \lya\ and LyC in LCEs either emerge directly from the central starbursts or escape isotropically in all directions. Strong LCEs, hosting a highly ionized ISM with lower HI columns \citep{SaldanaLopez2022, Flury2025}, also show more compact and {\asl less luminous \lya\ halos} in emission, with shorter \rsLya\ and low HFs. Hereby, {\ASL a fraction of the} LyC photons will escape the ISM without being absorbed, {\ASL while the \lya\ radiation will transfer through the CGM with minimal resonant scattering.}
\end{itemize}

Despite the caveats described in this work, and the scatter in the underlying relations, LAHs stand as a valuable tool for estimating the contribution of galaxies to the ionizing budget, particularly during the EoR, where indirect methods for \fesclyc\ are the only option. The rapid increase in \lya\ observations with JWST is expanding our understanding on the role of star-forming galaxies in early structure formation and IGM evolution, though current studies rely on integrated spectra, therefore missing crucial spatial information at CGM scales. Looking ahead, and based on the outcome of this study, we encourage the community to push for NIRSpec/IFU observations of LAHs of distant galaxies \citep[e.g.,][]{Bunker2023}. 

\begin{acknowledgments}
The authors thank the anonymous referee for providing useful comments, which have certainly improved the quality of this paper. This research is based on observations made with the NASA/ESA Hubble Space Telescope obtained from the Space Telescope Science Institute, which is operated by the Association of Universities for Research in Astronomy, Inc., under NASA contract NAS 5–26555. These observations are from HST GO programs 17069, 14131, and 11107. A.S.L. acknowledges support from the Knut and Alice Wallenberg Foundation. M.J.H. is supported by the Swedish Research Council (Vetenskapsrådet) and is fellow of the Knut and Alice Wallenberg Foundation. A.L.R. acknowledges support from HST GO17069. F.L. acknowledges funding from the European Union’s Horizon 2020 research and innovation program under the Marie Sklodowska-Curie grant agreement No. C3UBES-101107619. A.L.R., M.S.O., and L.K. acknowledge support from HST GO-17069. R.A. acknowledges support of grant PID2023-147386NB-I00 funded by MICIU/AEI/10.13039/501100011033 and by ERDF/EU, and the Severo Ochoa grant CEX2021-001131-S. 
\end{acknowledgments}

%

\vspace{5mm}
\facilities{HST (COS, ACS/SBC and WFC/UVIS)}


\software{astropy \citep{astropyI, astropyII}, 
          linmix \citep{linmix},
          numpy \citep{numpy}, 
          photutils \citep{photutils}, 
          pysersic \citep{pysersic}, 
          scipy \citep{scipy}, 
          }



\appendix
\section{Data tables}\label{sec:appA}
In Table \ref{tab:LyaUV_analysis}, we list the different size measurements for both the UV and \lya\ emission, together with other archival properties such as redshifts and escape fractions \citep[\fesclyc,][]{Flury2022a}. Table \ref{tab:LAH_analysis} presents the LAH scale lengths and halo fractions derived from our morphological decomposition of the LaCOS LAHs.

\begin{table*}
\begin{center}
\caption{{\aref Sample properties and circularized UV and \lya\ sizes for LaCOS galaxies}.}
\begin{tabular}{ccccccccc}
\toprule
ObjectID & $z$ & $f_{\rm esc}^{\rm LyC}{\rm (COS)}$ & $r_{20}^{\rm UV}{\rm ~(kpc)}$ & $r_{50}^{\rm UV}{\rm ~(kpc)}$ & $r_{90}^{\rm UV}{\rm ~(kpc)}$ & $r_{20}^{\rm \lya}{\rm ~(kpc)}$ & $r_{50}^{\rm \lya}{\rm ~(kpc)}$ & $r_{90}^{\rm \lya}{\rm ~(kpc)}$ \\
\midrule
J011309 & $0.3062$ & $0.022_{-0.012}^{+0.016}$ & $0.34~\pm~0.01$ & $0.78~\pm~0.03$ & $3.51~\pm~0.53$ & $0.86~\pm~0.10$ & $2.81~\pm~0.33$ & $15.99~\pm~3.70$ \\
J012910 & $0.2800$ & $\leq 0.007$ & $0.36~\pm~0.01$ & $1.06~\pm~0.06$ & $5.65~\pm~1.88$ & $0.69~\pm~0.05$ & $2.63~\pm~0.19$ & $14.47~\pm~2.82$ \\
J072326 & $0.2969$ & $\leq 0.004$ & $0.28~\pm~0.01$ & $0.72~\pm~0.05$ & $3.86~\pm~1.33$ & $0.61~\pm~0.09$ & $2.18~\pm~0.20$ & $9.68~\pm~5.01$ \\
J081409 & $0.2272$ & $\leq 0.007$ & $0.66~\pm~0.02$ & $2.25~\pm~0.05$ & $16.28~\pm~1.29$ & $-$ & $-$ & $-$ \\
J082652 & $0.2972$ & $\leq 0.009$ & $0.31~\pm~0.03$ & $0.73~\pm~0.11$ & $3.13~\pm~2.86$ & $1.17~\pm~0.37$ & $3.70~\pm~0.88$ & $15.17~\pm~6.16$ \\
J090918 & $0.2816$ & $0.491_{-0.230}^{+0.417}$ & $0.20~\pm~0.02$ & $0.47~\pm~0.06$ & $4.57~\pm~4.76$ & $0.32~\pm~0.02$ & $0.98~\pm~0.15$ & $9.37~\pm~4.63$ \\
J091113 & $0.2622$ & $0.023_{-0.007}^{+0.018}$ & $0.26~\pm~0.01$ & $0.67~\pm~0.04$ & $4.24~\pm~1.29$ & $0.54~\pm~0.05$ & $1.89~\pm~0.14$ & $7.93~\pm~2.65$ \\
J091207 & $0.2470$ & $\leq 0.008$ & $0.61~\pm~0.05$ & $1.76~\pm~0.23$ & $18.20~\pm~3.55$ & $0.70~\pm~0.09$ & $1.94~\pm~0.41$ & $9.36~\pm~4.03$ \\
J091703 & $0.3004$ & $0.161_{-0.055}^{+0.073}$ & $0.22~\pm~0.01$ & $0.52~\pm~0.01$ & $3.34~\pm~0.38$ & $0.44~\pm~0.03$ & $2.26~\pm~0.33$ & $13.66~\pm~2.36$ \\
J092532 & $0.3013$ & $0.092_{-0.034}^{+0.019}$ & $0.26~\pm~0.01$ & $0.58~\pm~0.03$ & $3.92~\pm~1.02$ & $0.40~\pm~0.02$ & $1.08~\pm~0.07$ & $5.90~\pm~1.20$ \\
J092552 & $0.3142$ & $\leq 0.004$ & $0.61~\pm~0.06$ & $1.47~\pm~0.11$ & $13.34~\pm~7.75$ & $1.04~\pm~0.30$ & $2.50~\pm~1.16$ & $6.30~\pm~3.53$ \\
J093355 & $0.2913$ & $0.266_{-0.110}^{+0.106}$ & $0.27~\pm~0.02$ & $0.58~\pm~0.08$ & $4.45~\pm~3.60$ & $0.39~\pm~0.01$ & $1.22~\pm~0.10$ & $8.09~\pm~1.42$ \\
J095236 & $0.3187$ & $0.042_{-0.013}^{+0.021}$ & $0.52~\pm~0.02$ & $1.07~\pm~0.04$ & $3.51~\pm~0.60$ & $0.80~\pm~0.09$ & $2.29~\pm~0.50$ & $12.41~\pm~6.15$ \\
J095700 & $0.2444$ & $\leq 0.001$ & $1.41~\pm~0.04$ & $8.28~\pm~0.14$ & $20.54~\pm~0.13$ & $-$ & $-$ & $-$ \\
J095838 & $0.3017$ & $0.019_{-0.011}^{+0.028}$ & $0.31~\pm~0.02$ & $0.74~\pm~0.09$ & $2.80~\pm~0.92$ & $0.54~\pm~0.06$ & $1.66~\pm~0.23$ & $8.80~\pm~4.90$ \\
J105331 & $0.2526$ & $0.012_{-0.004}^{+0.006}$ & $0.30~\pm~0.00$ & $0.80~\pm~0.03$ & $4.01~\pm~0.36$ & $1.56~\pm~0.34$ & $4.67~\pm~0.42$ & $14.87~\pm~2.51$ \\
J110452 & $0.2801$ & $\leq 0.011$ & $0.40~\pm~0.01$ & $0.88~\pm~0.05$ & $4.69~\pm~1.70$ & $0.89~\pm~0.09$ & $2.82~\pm~0.27$ & $9.20~\pm~2.52$ \\
J112224 & $0.3048$ & $0.026_{-0.018}^{+0.056}$ & $0.24~\pm~0.03$ & $0.51~\pm~0.08$ & $2.78~\pm~3.57$ & $0.46~\pm~0.04$ & $1.45~\pm~0.24$ & $13.85~\pm~5.11$ \\
J113304 & $0.2414$ & $0.022_{-0.009}^{+0.022}$ & $0.40~\pm~0.01$ & $1.02~\pm~0.05$ & $7.03~\pm~1.83$ & $0.92~\pm~0.07$ & $3.24~\pm~0.22$ & $12.91~\pm~2.77$ \\
J115855 & $0.2430$ & $0.066_{-0.015}^{+0.030}$ & $0.27~\pm~0.00$ & $0.74~\pm~0.02$ & $4.90~\pm~0.49$ & $0.35~\pm~0.01$ & $1.08~\pm~0.07$ & $6.26~\pm~0.76$ \\
J115959 & $0.2679$ & $0.043_{-0.016}^{+0.067}$ & $0.31~\pm~0.02$ & $0.89~\pm~0.13$ & $7.82~\pm~4.55$ & $0.62~\pm~0.04$ & $2.10~\pm~0.15$ & $6.81~\pm~1.45$ \\
J120934 & $0.2193$ & $\leq 0.013$ & $0.21~\pm~0.00$ & $0.52~\pm~0.02$ & $2.63~\pm~0.17$ & $0.73~\pm~0.06$ & $2.50~\pm~0.12$ & $10.38~\pm~1.43$ \\
J121915 & $0.3038$ & $0.013_{-0.005}^{+0.016}$ & $0.43~\pm~0.04$ & $1.09~\pm~0.19$ & $9.24~\pm~6.80$ & $1.27~\pm~0.38$ & $7.71~\pm~2.63$ & $23.69~\pm~2.67$ \\
J124033 & $0.2834$ & $\leq 0.011$ & $0.31~\pm~0.01$ & $0.72~\pm~0.06$ & $11.16~\pm~6.64$ & $1.56~\pm~0.31$ & $3.49~\pm~0.64$ & $10.19~\pm~4.87$ \\
J124423 & $0.2394$ & $\leq 0.015$ & $0.50~\pm~0.02$ & $1.32~\pm~0.04$ & $4.57~\pm~0.64$ & $1.57~\pm~0.06$ & $3.92~\pm~0.18$ & $13.92~\pm~1.53$ \\
J124835 & $0.2634$ & $0.047_{-0.026}^{+0.043}$ & $0.32~\pm~0.00$ & $0.75~\pm~0.01$ & $4.16~\pm~0.36$ & $0.66~\pm~0.01$ & $1.87~\pm~0.03$ & $7.06~\pm~0.34$ \\
J125503 & $0.3119$ & $\leq 0.009$ & $0.48~\pm~0.02$ & $1.04~\pm~0.04$ & $3.06~\pm~0.46$ & $0.84~\pm~0.06$ & $2.24~\pm~0.24$ & $15.08~\pm~3.97$ \\
J125718 & $0.3131$ & $\leq 0.014$ & $0.32~\pm~0.02$ & $0.92~\pm~0.10$ & $11.38~\pm~4.96$ & $0.62~\pm~0.08$ & $1.54~\pm~0.20$ & $5.39~\pm~2.69$ \\
J130559 & $0.3157$ & $0.178_{-0.058}^{+0.078}$ & $0.20~\pm~0.03$ & $0.52~\pm~0.12$ & $5.19~\pm~6.39$ & $0.26~\pm~0.02$ & $0.61~\pm~0.11$ & $6.05~\pm~6.23$ \\
J131037 & $0.2831$ & $0.016_{-0.006}^{+0.020}$ & $0.26~\pm~0.01$ & $0.64~\pm~0.04$ & $3.91~\pm~1.73$ & $0.45~\pm~0.04$ & $1.37~\pm~0.24$ & $8.73~\pm~2.63$ \\
J131419 & $0.2961$ & $\leq 0.001$ & $0.89~\pm~0.03$ & $2.02~\pm~0.04$ & $6.00~\pm~0.89$ & $1.63~\pm~0.36$ & $3.83~\pm~0.94$ & $13.54~\pm~4.32$ \\
J131904 & $0.3176$ & $\leq 0.002$ & $0.52~\pm~0.04$ & $1.34~\pm~0.15$ & $11.35~\pm~5.47$ & $0.51~\pm~0.14$ & $1.29~\pm~0.65$ & $6.76~\pm~7.55$ \\
J132633 & $0.3177$ & $0.118_{-0.084}^{+0.137}$ & $0.29~\pm~0.01$ & $0.75~\pm~0.05$ & $4.11~\pm~0.97$ & $0.46~\pm~0.03$ & $1.76~\pm~0.25$ & $12.79~\pm~3.85$ \\
J132937 & $0.3091$ & $\leq 0.001$ & $1.52~\pm~0.03$ & $3.18~\pm~0.05$ & $9.52~\pm~1.03$ & $1.84~\pm~0.31$ & $3.30~\pm~1.22$ & $12.20~\pm~7.28$ \\
J134559 & $0.2373$ & $\leq 0.002$ & $0.78~\pm~0.02$ & $1.68~\pm~0.04$ & $7.77~\pm~0.42$ & $-$ & $-$ & $-$ \\
J140333 & $0.2816$ & $0.031_{-0.014}^{+0.019}$ & $0.35~\pm~0.03$ & $1.20~\pm~0.33$ & $17.77~\pm~2.66$ & $0.40~\pm~0.05$ & $0.99~\pm~0.18$ & $3.61~\pm~1.97$ \\
J144010 & $0.3008$ & $0.005_{-0.002}^{+0.002}$ & $0.32~\pm~0.01$ & $0.85~\pm~0.02$ & $5.05~\pm~0.33$ & $0.54~\pm~0.04$ & $2.17~\pm~0.22$ & $13.11~\pm~1.88$ \\
J154050 & $0.2944$ & $\leq 0.001$ & $0.45~\pm~0.01$ & $1.11~\pm~0.02$ & $5.18~\pm~0.22$ & $1.46~\pm~0.37$ & $4.27~\pm~0.61$ & $14.02~\pm~2.40$ \\
J155945 & $0.2268$ & $\leq 0.025$ & $0.26~\pm~0.01$ & $0.69~\pm~0.03$ & $7.25~\pm~2.23$ & $0.68~\pm~0.08$ & $2.21~\pm~0.16$ & $6.13~\pm~0.83$ \\
J160437 & $0.3123$ & $\leq 0.007$ & $0.34~\pm~0.02$ & $0.78~\pm~0.08$ & $3.50~\pm~2.79$ & $1.08~\pm~0.09$ & $3.00~\pm~0.44$ & $23.36~\pm~3.78$ \\
J164607 & $0.2906$ & $0.023_{-0.010}^{+0.010}$ & $0.26~\pm~0.01$ & $0.59~\pm~0.04$ & $2.85~\pm~0.85$ & $0.68~\pm~0.06$ & $2.17~\pm~0.20$ & $10.58~\pm~2.85$ \\
J172010 & $0.2938$ & $0.031_{-0.014}^{+0.026}$ & $0.58~\pm~0.03$ & $1.26~\pm~0.06$ & $3.08~\pm~0.58$ & $0.83~\pm~0.07$ & $2.06~\pm~0.15$ & $9.44~\pm~3.06$ \\
\bottomrule
\end{tabular}
\end{center}
\label{tab:LyaUV_analysis}
{\bf Notes.} Column 1: object identifier. Column 2: spectroscopic redshift (from SDSS). Column 3: absolute ionizing escape fraction \citep{Flury2022a}. Column 4, 5 and 6: UV 20\%, 50\% (half-light) and 90\%-light radii (in kpc) from the UV continuum (F165LP) images \citep{LeReste2025}. Column 7, 8 and 9: \lya\ 20\%, 50\% and 90\%-light radii (in kpc) for the \lya\ emission, measured from a curve-of-growth analysis over the \lya\ images. 
\end{table*}

\begin{table*}
\begin{center}
\caption{{\aref Best-fit morphological parameters of the extended LAHs in LaCOS.}}
\begin{tabular}{cccc}
\toprule
ObjectID & $r_{s}^{\rm UV}{\rm ~(kpc)}$ & $r_{s}^{\rm \lya}{\rm ~(kpc)}$ & HF \\
\midrule
J011309 & $0.38~\pm~0.02$ & $5.02~\pm~0.60$ & $0.45~\pm~0.04$ \\
J012910 & $0.34~\pm~0.03$ & $4.94~\pm~0.45$ & $0.44~\pm~0.03$ \\
J072326 & $0.17~\pm~0.02$ & $2.82~\pm~0.26$ & $0.35~\pm~0.04$ \\
J081409 & $-$ & $-$ & $-$ \\
J082652 & $0.44~\pm~0.04$ & $6.93~\pm~0.89$ & $0.20~\pm~0.05$ \\
J090918 & $0.10~\pm~0.02$ & $1.58~\pm~0.68$ & $0.72~\pm~0.04$ \\
J091113 & $0.17~\pm~0.02$ & $2.30~\pm~0.20$ & $0.42~\pm~0.04$ \\
J091207 & $0.72~\pm~0.03$ & $3.83~\pm~0.57$ & $0.50~\pm~0.05$ \\
J091703 & $0.14~\pm~0.02$ & $5.41~\pm~0.47$ & $0.51~\pm~0.03$ \\
J092532 & $0.23~\pm~0.02$ & $1.48~\pm~0.22$ & $0.66~\pm~0.04$ \\
J092552 & $0.72~\pm~0.03$ & $7.23~\pm~0.82$ & $0.17~\pm~0.05$ \\
J093355 & $0.24~\pm~0.03$ & $2.83~\pm~0.38$ & $0.74~\pm~0.02$ \\
J095236 & $0.66~\pm~0.03$ & $4.68~\pm~0.56$ & $0.42~\pm~0.05$ \\
J095700 & $-$ & $-$ & $-$ \\
J095838 & $0.30~\pm~0.03$ & $2.41~\pm~0.37$ & $0.50~\pm~0.05$ \\
J105331 & $0.25~\pm~0.02$ & $6.47~\pm~0.50$ & $0.13~\pm~0.05$ \\
J110452 & $0.42~\pm~0.02$ & $4.33~\pm~0.37$ & $0.29~\pm~0.04$ \\
J112224 & $0.23~\pm~0.04$ & $0.93~\pm~0.13$ & $0.46~\pm~0.08$ \\
J113304 & $0.49~\pm~0.02$ & $4.82~\pm~0.28$ & $0.27~\pm~0.02$ \\
J115855 & $0.19~\pm~0.01$ & $4.11~\pm~0.30$ & $0.66~\pm~0.02$ \\
J115959 & $0.38~\pm~0.06$ & $2.91~\pm~0.23$ & $0.37~\pm~0.03$ \\
J120934 & $0.12~\pm~0.02$ & $4.14~\pm~0.27$ & $0.38~\pm~0.03$ \\
J121915 & $0.61~\pm~0.04$ & $2.91~\pm~0.48$ & $0.44~\pm~0.08$ \\
J124033 & $0.30~\pm~0.02$ & $5.04~\pm~0.54$ & $0.12~\pm~0.05$ \\
J124423 & $0.76~\pm~0.04$ & $5.24~\pm~0.31$ & $0.22~\pm~0.02$ \\
J124835 & $0.41~\pm~0.01$ & $2.24~\pm~0.06$ & $0.41~\pm~0.02$ \\
J125503 & $0.64~\pm~0.02$ & $3.25~\pm~0.44$ & $0.46~\pm~0.04$ \\
J125718 & $0.19~\pm~0.02$ & $1.51~\pm~0.15$ & $0.23~\pm~0.06$ \\
J130559 & $0.13~\pm~0.03$ & $1.64~\pm~0.48$ & $0.90~\pm~0.05$ \\
J131037 & $0.22~\pm~0.03$ & $4.50~\pm~0.65$ & $0.59~\pm~0.04$ \\
J131419 & $1.05~\pm~0.02$ & $7.61~\pm~0.82$ & $0.19~\pm~0.06$ \\
J131904 & $0.56~\pm~0.04$ & $2.35~\pm~0.51$ & $0.50~\pm~0.12$ \\
J132633 & $0.24~\pm~0.03$ & $2.76~\pm~0.43$ & $0.52~\pm~0.05$ \\
J132937 & $1.97~\pm~0.05$ & $5.55~\pm~0.65$ & $0.12~\pm~0.10$ \\
J134559 & $-$ & $-$ & $-$ \\
J140333 & $0.25~\pm~0.03$ & $1.39~\pm~0.33$ & $0.48~\pm~0.08$ \\
J144010 & $0.23~\pm~0.02$ & $6.74~\pm~0.42$ & $0.54~\pm~0.03$ \\
J154050 & $0.61~\pm~0.02$ & $7.55~\pm~0.59$ & $0.17~\pm~0.05$ \\
J155945 & $0.29~\pm~0.02$ & $3.10~\pm~0.19$ & $0.25~\pm~0.03$ \\
J160437 & $0.47~\pm~0.05$ & $2.39~\pm~0.33$ & $0.28~\pm~0.06$ \\
J164607 & $0.22~\pm~0.02$ & $3.67~\pm~0.52$ & $0.46~\pm~0.04$ \\
J172010 & $0.87~\pm~0.04$ & $4.98~\pm~0.61$ & $0.50~\pm~0.03$ \\
\bottomrule
\end{tabular}
\end{center}
\label{tab:LAH_analysis}
{\bf Notes.} Column 1: object identifier. Columns 2 and 3: \lya\ core (S{\'e}rsic) and halo (exponential) scale lengths (in kpc), from our 2D modeling to the observed LAHs. Columns 4: \lya\ halo fraction. 
\end{table*}


\newpage
\bibliography{LaCOS_LAHs}{}
\bibliographystyle{aasjournal}



\end{document}